\documentclass[10pt]{article}

\usepackage{amsmath,amsfonts,amssymb,epsfig} %

\usepackage[normalem]{ulem}
\usepackage{graphicx}
\usepackage[arrow, matrix, curve]{xy}
\usepackage{txfonts}
\usepackage[top=1in, bottom=1in, left=1in, right=1in]{geometry}
\usepackage[usenames,dvipsnames]{color}

\usepackage{hyperref}

\def\define{\triangleq}

\def\R{\mathbb{R}}

\def\Rplus{\mathbb{R}^{++}}
\def\Rpplus{\mathbb{R}^{++}}

\def\U{{U}}
\def\Util{{}}

\def\P{\mathbb{P}}
\def\E{\mathbb{E}}

\def\ind{\mathbb{I}}

\def \NT {\mathbf{NT^{(j)}}}
\def \NTw {\mathbf{NT^{(w)}}}
\def \NTo {\mathbf{NT^{(1)}}}
\def \NTY {\mathbf{NT_Y^{(j)}}}
\def \NTYw {\mathbf{NT_Y^{(w)}}}

\def \B{\mathbf{B^{(j)}}}

\def \S{\mathbf{S^{(j)}}}
\def \K{\mathbb{K}}

\def \Qjpm {Q^{(j)^\pm}}

\def \jpm {{ (j)^{\pm}}}
\def \jp {{ (j)^{+}}}
\def \jm {{ (j)^{-}}}
\def \jstr {{ (j)^{*}}}
\def \j {{ (j)}}
\def \one {^{(1)}}
\def \ones {^{(1)^{*}}}

\def \onepm {^{(1)^{\pm}}}
\def \w {^{(w)}}
\def \ws {^{(w)^{*}}}
\def \wpm {^{(w)^{\pm}}}
\def\ju {^{\j}}
\def \yjstr {y^{\jstr}}

\def \y {{\tilde y}}
\def \H {{\tilde H}}
\def \zeroClosedTopen {{ [0,T)}}
\def \zeroClosedTclosed {{ [0,T]}}
\def \tClosedTclosed {{ [ t,T] }}

\def \tClosedtOneClosed {{ [t,t_1]}}

\def \dom{{ \R\times\R\times\Rpplus}}

\def \D{\mathcal{D}}

\newcommand{\e}[1]{\operatorname{e}^{#1}}

\def \abs#1{ \left \vert #1 \right \vert }
\def \sign#1{\operatorname{sign} \left\{#1 \right\} }

\def \Ind {{\ind_{\{j=w\}}}}

\newcommand{\du}{\mathrm{d}}

\newtheorem{theorem}{Theorem}[section]
\newtheorem{assumption}[theorem]{Assumption}

\newtheorem{definition}[theorem]{Definition}
\newtheorem{lemma}[theorem]{Lemma}

\newtheorem{Cor}[theorem]{Corollary}

\newtheorem{Comparison Theorem}[theorem]{Comparison Theorem}
\newtheorem{remark}[theorem]{Remark}
\newtheorem{example}[theorem]{Example}

\title{Pricing a Contingent Claim Liability with Transaction Costs Using Asymptotic Analysis for Optimal
Investment   }
\date{\today}

\author{Maxim Bichuch%
\footnote{The author acknowledge partial financial supported by NSF grant DMS-0739195. The author wants to thank Paolo Guasoni, Johannes Muhle-Karbe, Steven Shreve and Stephan Sturm for helpful discussions.}\\
Department of Operations Research \& Financial Engineering\\
Princeton University\\
Princeton, NJ 08544\\
mbichuch@princeton.edu\\
}

\begin{document}

\maketitle

\begin{abstract}
We price a contingent claim liability using the utility indifference argument. We consider an agent with exponential utility, who invests in a stock and a money market account with the goal of maximizing the utility of his investment at the final time $T$ in the presence of a proportional transaction cost $\varepsilon>0$ in two cases with and without a contingent claim liability. Using the computations from the heuristic argument in Whalley \&  Wilmott \cite{WhalleyWilmott} we provide a rigorous derivation of the asymptotic expansion of the value function in powers of $\varepsilon^{1/3}$ in both cases with and without a contingent claim liability. Additionally, using utility indifference method we derive the price of the contingent claim liability up to order $\varepsilon$. In both cases, we also obtain a ``nearly optimal" strategy, whose expected utility asymptotically matches the leading terms of the value function.
\end{abstract}




\setcounter{theorem}{0}
\setcounter{equation}{0}

\section{Introduction}
In a complete, frictionless market every contingent claim with maturity time $T$ can be replicated. However, in the presence of proportional transaction costs, even in the simple case of a call option this replication becomes prohibitively expensive, since the hedging portfolio cannot be continuously rebalanced. Thus we are forced to search for an alternative approach. Instead of a perfect hedge, one can look for a dominating portfolio. In that case, the price of the option by simple arbitrage argument cannot be greater then the initial price of the dominating portfolio. A trivial portfolio that dominates a call option is just holding one share of stock. It turns out that in the presence of transaction costs this is also the cheapest dominating portfolio. See \cite{BouchardTouzi}, \cite{CvitanicPhamTouzi}, \cite{DelbaenKabanovValkeila}, %
\cite{KabanovLast}, \cite{KabanovStricker}, \cite{KoehlPhamTouzi97}, \cite{KoehlPhamTouzi98}, \cite{LeventhalSkorohod}, \cite{SonerShreveCvitanic}, so a different approach is required. 

A natural alternative to continuous hedging, is to allow only trades in discrete time. Leland \cite{Leland} follows this approach and considers a discrete time model with transaction costs. Using a formal delta hedge argument, he derives a price for an option by modifying the volatility in the Black-Scholes formula. This result was made rigorous by Lott \cite{Lott}. Similarly, Boyle \& Vorst \cite{BoyleVorst} in a a binomial stock model showed that as the time step and the transaction costs go to zero at the appropriate rate, the option price converges to a  Black-Scholes price with an adjusted volatility.

An alternative approach using preferences %
was proposed by Hodges \& Neuberger \cite{HodgesNeuberger}. Their idea is to price an option so that a utility maximizer is indifferent between either having a certain initial capital for investment or else holding the option but having initial capital reduced by the price of the option. This constitutes an extension to the Black-Scholes hedging approach, since in a frictionless market, the option price being a result from this utility maximization problem, matches the Black-Scholes price. Dumas and Luciano \cite{DumasLuciano} also make this point. This utility-based option pricing has received a lot of attention since then. To name a few: Bouchard \cite{Bouchard} defined a dual problem to the utility-based option pricing problem, and proved that it admits a solution and characterized the optimal portfolio process in terms of the solution of the dual problem. Clewlow \& Hodges \cite{ClewlowHodges} numerically computed the optimal strategy for the utility indifference hedge. Constantinides \& Zariphopoulou \cite{ConstantinidesZariphopoulou99} derived an upper bound on the price of a call option under some restrictions on the utility function, and later in \cite{ConstantinidesZariphopoulou01} also derived both upper and lower bounds of price of derivatives in case of constant relative risk aversion utility.
Furthermore, Bouchard, Kabanov \& Touzi \cite{BouchardKabanovTouzi} showed that if the seller is strongly risk-averse, his utility indifferencce price approaches the super-replication price increased by the liquidation value of the initial endowment.

This approach is further utilized in celebrated papers by Davis, Panas \& Zariphopoulou \cite{DavisPanasZariphopoulou} and  Whalley \& Wilmott \cite{WhalleyWilmott}, who considered the problem of pricing an European option in a market with proportional transaction costs $\varepsilon>0$. The former showed that to find the option price one needs to solve two stochastic optimal control problems. They also showed that the value functions of these problems are the unique viscosity solutions, with different boundary conditions, of a fully nonlinear quasi-variational inequality. Whalley \& Wilmott \cite{WhalleyWilmott} used these results to formally derive an asymptotic power expansion of the value functions, in powers of $\varepsilon^{\frac13}$. They start by assuming a power expansion for these value functions and by formally matching coefficients, they computed the leading terms in both the case of holding the option liability and the case without it. Barles \& Soner \cite{BarlesSoner} performed an alternative asymptotic analysis in the same model assuming that both the transaction costs and the hedgerÕs risk tolerance are small. They showed that the option price is the unique solution to a nonlinear Black-Scholes equation with an adjusted volatility.
The starting point of this paper is to consider the two optimal investment problems with and without contingent claim liability in finite time in a market with proportional transaction costs. There is an extensive literature on the later problem starting from Merton \cite{Merton} who was the first to consider the problem of optimal investment and consumption in infinite time horizon without transaction costs, and found that the agent's optimal strategy is to keep a constant proportion of wealth, invested in stock. Transaction costs were introduced into Merton's model by  Magill \& Constantinides \cite{MagillConstantinides}. Their analysis of the infinite time horizon investment and consumption problem, despite being heuristic, gives an insight into the optimal strategy and the existence of the {\em``no-trade"} region. A more rigorous analysis, though under restrictive conditions, of the same infinite time horizon problem was given by Davis \& Norman \cite{DavisNorman}. The viscosity solution approach to that infinite time horizon problem was pioneered by Shreve \& Soner \cite{ShreveSoner}, who also showed that the value function is smooth.

When $\varepsilon>0$, the optimal policy is to trade as soon as the position is sufficiently far away from the Merton proportion. More specifically, the agent's optimal policy is to maintain her position inside a region that we refer to as the {\em``no-trade"} (NT) region. If the agent's position is initially outside the NT region, she should immediately sell or buy stock in order to move to its boundary. The agent then will trade only when her position is on the boundary of the NT region, and only as much as necessary to keep it from exiting the NT region, while no trading occurs in the interior of the region; see Davis, Panas \& Zariphopoulou \cite{DavisPanasZariphopoulou}. There is a trade-off between the amount of transaction costs payed due to portfolio rebalancing and the width of the NT region. Reducing the hedging error generally increases the transaction costs, and vice versa. Not surprisingly, the width of the NT region depends on time, which makes it difficult to pinpoint exactly the optimal policy. A useful and perhaps more informative approach for obtaining explicit results, the approach of this paper, is to develop a power series expansion for the value function and the boundaries of the NT region in powers of $\varepsilon^\frac13$. This approach was pioneered by Jane\v{c}ek \& Shreve \cite{JanecekShreve1} in solving an infinite horizon investment and consumption problem. It is also used in Bichuch \& Shreve \cite{BichuchShreve} in an infinite horizon investment and consumption problem problem with multiple risky assets, and in Bichuch \cite{Bichuch} to solve a finite horizon optimal investment problem. This approach also allows us to find a simple {\em``nearly-optimal"} policy, that matches the leading terms of the power expansion of the value function.

Our goal is to price the contingent claim liability using the utility indifference argument, in case of exponential utility. To achieve it we consider the problem of an agent seeking to optimally invest in the presence of proportional transaction costs in two cases with and without a contingent claim liability. The agent can invest in a stock, modeled as a geometric Brownian motion with drift $\mu$ and volatility $\sigma$, and in a money market with constant interest rate $r$. The agent pays proportional transaction cost $\varepsilon$ for trading stocks, with the goal of optimizing the total utility of wealth at the final time $T$, when she would be required to close out her stock position and pay the resulting transaction costs, and pay the liability in case she has one. We refer to this optimized utility of wealth as the value function. %
In this paper, we compute the asymptotic expansion of the value function up to and including the order $\varepsilon^\frac23.$ We also find a simple  {\em``nearly-optimal"} trading policy that, if followed, produces an expected utility of the final wealth that asymptotically matches the value function at the order of $\varepsilon^\frac23.$ As a corollary from utility indifference pricing we find the asymptotic expansion for the price of the contingent claim up and including the order $\varepsilon^\frac23.$

The rest of this paper is structured as follows: in Section \ref{sec:setup} of this paper we define our model, state the Hamilton-Jacobi-Bellman (HJB) equation, and define the problem in reduced variable. The main results of this paper are stated in Section \ref{sec:mainResults}. In Section \ref{sec:discussion} we consider the application of the main results to pricing European call and put options. In Section \ref{sec:expansion} based on results of the heuristic arguments of Whalley \&  Wilmott \cite{WhalleyWilmott} we define auxiliary functions and prove some preliminary results. We use these results to construct smooth functions $Q^{\jpm}$ for both cases with ($j=w$) and without ($j=1$) a contingent claim liability. In Section \ref{proofMainThm} we show that $Q^{\jpm}$ are sub- and supersolutions for the HJB equation and calculate the final time boundary conditions. We prove the existence of a ``nearly-optimal" in Section \ref{sec:nearly-optimal_strat} and prove the verification argument in Section \ref{sec:comp}. Finally, the proofs of some of the technical lemmas are deferred to the Appendix.

\section{Model definition}\label{sec:setup}

In this section we give the model definition. We use the same model as in Davis, Panas \& Zariphopoulou \cite{DavisPanasZariphopoulou} and Whalley \&  Wilmott \cite{WhalleyWilmott}. To fix notation, we will denote $\Rpplus\define\{x>0\}$.

We consider the problem of investment on a finite time interval $\zeroClosedTclosed$. Let $\{Z_t,t
\geq 0\}$ be a standard Brownian motion on a filtered probability space $\bigl(\Omega,\mathcal F,\{\mathcal F_t\}_{0 \le t \le T},\P\bigr)$. The market model evolution for the wealth invested in bonds, number of shares of stock, and stock price  starting at time $t\in\zeroClosedTclosed$ from $(B,y,S)\in\dom$ is given respectively for $s\in\tClosedTclosed$ by
\begin{eqnarray}
dB_s&=&rB_sds-(1+\varepsilon)S_sdL_s+(1-\varepsilon)S_sdM_s,~B_{t-}=B,\label{eq:dB}\\
dy_s&=&dL_s-dM_s,~y_{t-}=y\label{eq:dy}\\
dS_s&=&S_s(\mu ds+\sigma dZ_s),~S_{t}=S\label{eq:dS}
\end{eqnarray}
where $r, \mu, \sigma$ are positive constants, $\mu>r$, $\varepsilon\in(0,1)$ is the proportional transaction cost, $L_s,M_s$ are the cumulative number of shares bought or sold respectively by time $s$. The processes $B_s$ and $y_s$ start at time $t-$, as the agent may trade instantaneously and may rebalance his position at time $t$.

We define the cash value of a number of shares $y$ of stock $S$ as
\begin{equation}
c(y, S)\define(1-\varepsilon\sign{y})yS.
\label{eq:c}
\end{equation}
Throughout this paper, we will be considering two problems of optimal investment - the problem without a contingent claim liability, and the problem with European type contingent claim liability with smooth payoff function $g(S_T)$ at the final time $T$. These cases would be referred to as $j=1$ and $j=w$ respectively.

The wealth at the final time, without ($j=1$) and with the contingent claim liability ($j=w$) respectively is given by
\begin{eqnarray}
&&\Phi^{(1)}(T, B_T, y_T, S_T)\define B_T+c(y_T,S_T),\label{eq:Phi}\\
&&\Phi^{(w)}(T, B_T, y_T, S_T)\define B_T+c(y_T-g'(S_T),S_T) - (g(S_T)-g'(S_T)S_T).\nonumber
\end{eqnarray}
Note that in this case the contingent claim liability is settled in physical delivery. In this manuscript we will concentrate on exponential utility function 
\begin{equation}
\U (x)\define\Util-e^{-\gamma x},~x\in\R,
\label{eq:Util}
\end{equation}
with the parameter $\gamma>0$. 

For $(t, B, y, S) \in \zeroClosedTclosed\times\dom,$ we define the value function without and with the contingent claim liability respectively as
\begin{equation}
V^{(j)}(t,B,y,S)\define\sup%
\limits_{\mbox{admissible strategy}} \E_t^{B,y,S} \left[\U \left( \Phi^{(j)}(T,B_T, y_T,S_T)\right)\right], \mbox{ $j=1, w$},
\label{eq:V^j}
\end{equation}
where $\E_t^{B,y,S}[.]$ denotes the conditional expected value, conditioned on $B_{t-}=B,y_{t-}=y,S_{t}=S.$ %
We will withhold the definition of admissible strategies, until Section \ref{sec:mainResults}.
Clearly 
\begin{equation}
V^{(j)}(T,B,y,S)=\U\left(\Phi^{(j)}(T, B, y, S)\right),~j=1,w.
\label{eq:V-final-time}
\end{equation}
Define the second-order differential operator $\mathcal L$ as
\begin{equation} 
\mathcal L \psi\define -\psi_t-rB\psi_B-\mu S\psi_S-\frac12 \sigma^2S^2\psi_{SS}.
\label{eq:L}
\end{equation}
The following theorem  - Theorem 2 from Davis, Panas \& Zariphopoulou \cite {DavisPanasZariphopoulou}  is presented for completeness, but will not be used here. 

\begin{theorem} \label{thm:viscosol}
The value functions $V^{(j)}\in C\left(\zeroClosedTclosed\times\dom\right),~j=1,w$ are viscosity solutions of the Hamilton-Jacobi-Bellman (HJB) equation $\mathcal H_4 V^{(j)}=0$ on $\zeroClosedTclosed\times \dom$, where
\begin{equation}
\mathcal H_4 \psi \define\min \left\{-\psi_y+(1+\varepsilon)S\psi_B, \psi_y-(1-\varepsilon)S\psi_B, \mathcal L \psi\right\}.
\label{eq:HJB1-WW}
\end{equation}
This means that for fixed $j$ and for every point $(t,B,y,S)\in \zeroClosedTclosed\times\dom$ and for every $C^{1,1,1,2}$ function $\psi$ defined on $\zeroClosedTclosed\times\dom$, such that $\psi(t,B,y,S)=V^{(j)}(t,B,y,S)$ and $\psi\ge V^{(j)}$ (respectively $\psi\le V^{(j)}$) on $\zeroClosedTclosed\times\dom$ implies 
$$\min \left\{-\psi_y+(1+\varepsilon)S\psi_B, \psi_y-(1-\varepsilon)S\psi_B, \mathcal L \psi\right\}$$
is non-positive (respectively non-negative) at $(t,B,y,S)$.
\end{theorem}

\begin{remark}\label{remark:deriv}
Here and in the rest of this paper, when referring to the derivative with respect to $t$ at $t=0$ or $t=T$, we mean the appropriate one sided derivative.
\end{remark}

Let $\delta\define\delta(T,s)\define\e{-r(T-s)}$, for $j=1,w$ we define 
\begin{equation}
Q^{(j)}(S,y,s) \define \Util-V^{(j)}(s,0,y,S),~~(s, y,S)\in \zeroClosedTclosed\times\R\times\Rpplus.
\label{eq:Q}
\end{equation}
Davis, Panas \& Zariphopoulou \cite {DavisPanasZariphopoulou} show that the following representation holds
\begin{equation}
V^{(j)}(s,B,y,S)=\Util-\exp{ \left\{-\gamma \frac{B}{\delta(T,s)}\right\}}Q^{(j)}(S,y,s), ~(s, B, y,S)\in \zeroClosedTclosed\times\dom.
\label{eq:V-Q}
\end{equation}
Moreover, for $(y,S)\in\R\times\Rpplus$ from \eqref{eq:V-final-time} the final time conditions for $Q^{(j)}$ are
\begin{equation}
Q^{(j)}(S,y,T)=-\U ( \Phi^{(j)}(T,0, y,S))=\exp{\left\{-\gamma\left[c(y-g'(S)\Ind,S) - (g(S)-g'(S)S)\ind_{\{j=w\}}\right]\right\}},~j=1,w.\label{eq:Q-final-time}
\end{equation}
\begin{remark}\label{remark:viscsol-WW}
It follows that for $(t,B,y,S)\in\zeroClosedTclosed\times\dom$
\begin{equation}
\mathcal H_4V\ju(t,B,y,S)=\Util\exp{\left\{-\frac{\gamma B}{\delta} \right\} }\mathcal H_3Q\ju(S,y,t),~j=1,w,
\label{eq:HJB-WW1}
\end{equation}
where
\begin{equation}
\mathcal H_3 \psi \define \min\left\{\psi_y+\frac{(1+\varepsilon)\gamma S\psi}{\delta}, -\psi_y-\frac{(1-\varepsilon)\gamma S\psi}{\delta}, \D \psi\right\},
\label{eq:HJB-WW}
\end{equation}
and
\begin{equation}
\D \psi\define\psi_t+\mu S\psi_S+\frac12 \sigma^2S^2\psi_{SS}.
\label{eq:D}
\end{equation}
\end{remark}

\section{Main Results}
\setcounter{equation}{0} \label{sec:mainResults}
In this section we present the main theorems of this paper. The main idea of the paper, is to find tight upper and lower bounds on the functions $Q\ju$, and thus using \eqref{eq:V-Q} on $V\ju$, for both cases with and without contingent claim liability. These tight bounds would match at the order of $\varepsilon^\frac23$, which would allow us to identify $V\ju$ and to compute the utility indifference price of the contingent claim, both at the $O(\varepsilon^\frac23)$.

First, we present a brief overview of the utility optimization problem in case of zero transaction costs. In the rest of the section we present the main theorems. In Theorem \ref{thm:comp} we construct the functions $Q^{\jpm}$ and prove that they are tight lower and upper bounds on the functions $Q^{\j}$, $j=1,w$. The intuition behind the construction of $Q^{\jpm}$ is the heuristic computations appearing in Whalley \&  Wilmott \cite{WhalleyWilmott}. To build $Q^{\jpm}$ we will need to define the appropriate {\em``no-trade"} regions $\NT$ that would be centered at the optimal trading strategy for the case of zero transaction costs $y^{\jstr}$ and have the width of order $\varepsilon^\frac13$.
It can be easily shown that $Q^{\jpm}$ are viscosity sub- and supersolutions of the HJB equation $\mathcal H_3 Q^{\jpm}=0$; see \cite{CL}, \cite{CEL}, \cite{CrandallIshiiLions}. Moreover, using $Q^{\jpm}$ one could easily construct $V^{\jpm}$ to be viscosity sub- and supersolutions of the HJB equation $\mathcal H_4 V^{\jpm}=0$.

We have already stated the classical Theorem \ref{thm:viscosol} asserting that the value functions $V^{\j}$ are viscosity solutions of the HJB equation $\mathcal H_4 V^{\j}=0$. One way to proceed to establish that supersolutions and subsolutions are indeed upper and lower bounds of the value function is to use a comparison theorem. However, the problem in this approach is that $V^{\j}$ are discontinuous on the boundary $S=0$. Intuitively, whenever $S>0$ there are two assets to invest -- the money market and the stock, however, when $S=0$ there is only one risk free asset, and this creates the discountinuity. One can, of course, redefine $V^{\j}$ to be its limit on that boundary in order to make it continuous. This would not change the problem, since the probability of reaching that boundary is zero, but that creates a different problem of establishing the boundary conditions of $V^{\j}$, which is hard.

Hence we approach this problem from a different angle. We use a version of Verification Argument lemma from Stochastic Calculus, to establish the desired dominating property.

\begin{assumption}\label{A1}
We will assume that the contingent claim payoff function $g:[0,\infty)\rightarrow [0,\infty)$ is four times continuously differentiable, and such that $g(S)-g'(S)S, S^2g''(S)$, $S^3g^{(3)}(S)$ and $S^4g^{(4)}(S)$ are all bounded for $S\in[0,\infty)$. 
\end{assumption}

\begin{assumption}\label{A2}
We will also assume that $\frac{\e{ -rT } (\mu-r)}{\gamma\sigma^2}-\sup\limits_{S\in[0,\infty)}S^2\abs{g''(S)}\ge \varepsilon_1$, for some $\varepsilon_1>0$. The supremum above is finite by Assumption \ref{A1}.
\end{assumption}

For $(S,t)\in\Rpplus\times\zeroClosedTclosed$ let $V_0(S,t)$ be the Black-Scholes price of the contingent claim liability, i.e. $V_0$ satisfies 
\begin{equation}
V_{0_t} +\frac{\sigma^2S^2}{2}V_{0_{SS}} +rSV_{0_S}-rV_0=0,%
\label{eq:BS}
\end{equation}
with final time condition $V_0(S,T)=g(S).$
In order to define the value function in case of zero transaction costs, we will also need the following definitions (see also Remark \ref{remark:zero-trans-costs}). For $(S,t)\in \Rpplus\times\zeroClosedTclosed,~j=1,w$  let
\begin{eqnarray}
H_0^{(j)}(S,t)&\define&-\frac{(\mu-r)^2(T-t)}{2\sigma^2}+\frac{\gamma}{\delta}V_0(S,t)\ind_{\{j=w\}},\label{eq:H_0}\\
y^{\jstr}(S,t)&\define&\frac\delta\gamma H_{0_S}^{(j)}+\frac{\delta(\mu-r)}{\gamma S\sigma^2}.\label{eq:y-star}
\end{eqnarray}
Additionally, the following would be used to define the width of the NT region. For $(S,t)\in \Rpplus\times\zeroClosedTclosed$ and $j=1,w$ let
\begin{eqnarray}
Y^{\j}(S,t)&\define& \left(\frac{3S\delta \left(y_S^{\jstr}(S,t)\right)^2}{2\gamma} \right)^{\frac13},
\label{eq:Y-pm}\\
y^{\jpm}(S,t)&\define& y^{\jstr}(S,t) \pm \varepsilon^{\frac13} Y^{\j}(S,t).
\label{eq:y_Ypm}
\end{eqnarray}
\begin{remark}\label{remark:natation1}
For convenience, for $j=1,w$ we will drop the parameters and refer to $y^{\jstr}(S,t),Y^{\j}(S,t),y^{\jpm}(S,t)$ from \eqref{eq:y-star}, \eqref{eq:y_Ypm}, \eqref{eq:Y-pm} simply as $y^{\jstr},Y^{\j},y^{\jpm}$.
\end{remark}
Sicne $Y^{\j}\ge0,~j=1,w$  we can finally define the {\em``no-trade"} region as 
\begin{equation}
\NT=\Big\{(S,y,t)\in\Rpplus\times\R\times\zeroClosedTclosed \colon ~~\abs{y-y^{\jstr}} < \varepsilon^{\frac13}Y^{\j}\Big\},~~j=1,w.
\label{eq:NT}
\end{equation}

\begin{remark}\label{remark:zero-trans-costs}
It can be shown that in case of zero transaction, and in case of no contingent claim liability for
 $(t, B,y,S)\in\zeroClosedTclosed\times\R\times\R\times\Rpplus$ the value function
$$%
V^{\one}(t,By,S)= -\exp{ \left\{-\gamma \frac{B}{\delta}\right\}}\exp{\left\{  -\frac{\gamma}{\delta}Sy +H_0^{(1)}(S,t) \right\} },$$
and the optimal number of stocks is given by $y^{\ones}$; see e.g. Pham \cite{Pham}.
Moreover, since in the Black-Scholes model the market is complete, it follows from Davis, Panas \& Zariphopoulou \cite {DavisPanasZariphopoulou} that the value function
$$%
V^{\w}(t,B,y,S)= -\exp{ \left\{-\gamma \frac{B}{\delta}\right\}}\exp{\left\{  -\frac{\gamma}{\delta}Sy +H_0^{(w)}(S,t) \right\} },$$
and the optimal number of stocks is given by $y^{\ws}$; see again Pham \cite{Pham}. Note that in this case, with no transaction costs, there is no difference between physical and cash settlements.

With positive transaction costs it will be prohibitively expensive to adhere to this optimal strategy. Intuitively it would not be optimal to trade stocks, when the number of stocks differs from $y^{\jstr}$ by at most $Y^{\j}\varepsilon^\frac13$, i.e. in the {\em``no-trade"} region $\NT$ (see e.g. Whalley \& Wilmott \cite{WhalleyWilmott} ). In this paper, we will quantify this intuition.

\end{remark}

We can now rigorously define what is an admissible strategy:
\begin{definition}\label{def:admis}
For $j=1,w$, and $t\in\zeroClosedTclosed$, let $(L^{\j},M^{\j})$ be a trading strategy, and let the process $(s, B_s^{\j},y_s^{\j},S_s),$  $s\in\tClosedTclosed$ given by \eqref{eq:dB}--\eqref{eq:dS} starting at time $t$ from $(B,y,S)$. We say that a trading strategy $(L^{\j},M^{\j})$ is admissible, if it satisfies the following conditions:

\begin{enumerate}
\item $\left\{ \exp{ \left\{  -\gamma \left( B_\tau^{\j} + y_\tau^{\j} S_\tau - V_0(S_\tau,\tau)\ind_{\{j=w\}}  \right)\right\}  } \right\}\Big|_{\tau\in\mathcal T_T}$ is uniformly integrable, %
where $\mathcal T_T$ is the set of all stopping times $\tau\le T$a.s..
\item $\abs{y_s^{\j}-\yjstr}S_s,~s\in\tClosedTclosed$ is bounded. %
\end{enumerate}
We will denote the set of all such strategies by $\mathcal T^{(j)}(t,B,y,S)$. 
\end{definition}
\begin{lemma}\label{lemma:existence}
For $j=1,w$, fix the initial position at time $t\in\zeroClosedTclosed$ to be $(B,y,S)\in\R\times\R\times\Rpplus$, and let $(\tilde L^{\j},\tilde M^{\j})$ be the strategy  associated with the {\em``no-trade"} region $\NT$. Then there exists a strong solution $(s, \tilde B^{\j}_s, \y^{\j}_s, S_s)\big|_{s\in\tClosedTclosed}$ to \eqref{eq:dB} - \eqref{eq:dS}, starting at $(\tilde B^{\j}_{t-},\y^{\j}_{t-}, S_t)$ $=(B,y,S)$ such that $(S_s,\y^{\j},s)$ is a reflected process inside $\overline\NT$. Moreover, $\E_t^{B,y,S}\left[   \exp{ \left\{  -\gamma ( \tilde B^{\j}_T + \y^{\j}_TS_T -  V_0(S,T) \ind_{\{j=w\}}\right\}  } \right] $ is finite for $\varepsilon$ small enough, and the strategy $(\tilde L^{\j},\tilde M^{\j})$ is admissible.
\end{lemma}
The proof is deferred to Section \ref{sec:nearly-optimal_strat}.

Additionally, for $(S,t)\in \Rplus\times\zeroClosedTclosed$ define the functions $H_2^{(j)}(S,t),~j=1,w$ to be the solution of 
\begin{equation}
H_{2_t}^{(j)}+rSH_{2_S}^{(j)}+\frac{\sigma^2S^2}{2}H_{2_{SS}}^{(j)}=-\frac12\left(\frac{3\gamma^2S^4\sigma^3 \left(y_S^{\jstr}\right) ^2}  {2\delta^2} \right)^{\frac23},
\label{eq:H_2}
\end{equation}
with final time condition $H_2^{(j)}(S,T)=0$. The existence and uniqueness of $H_2^{\j}$ is shown in Lemma \ref{lemma:bounds}.

The main theorems of this paper are:

\begin{theorem}\label{thm:Thm2}
Assume $\gamma>0$, and assume that $g$, the payoff of a European style contingent claim with maturity $T>0$, satisfies Assumptions \ref{A1} and \ref{A2}. Fix a compact $\K\subset\R\times\Rpplus$. Then for $j=1,w$ and for $(t, B,y,S)\in\zeroClosedTclosed\times\R\times\K$ the value function for $\varepsilon>0$ small enough is

\begin{eqnarray}
&&V^{(j)}(t,B,y,S)=\sup\limits_{(L,M)\in\mathcal T^{\j}(t,B,y,S)} \E_t^{B,y,S} \left[\U ( \Phi^{(j)}(T,B_T, y_T,S_T))\right]\label{eq:V-expan}\\
&&=\Util-\exp{ \left\{-\gamma \frac{B}{\delta}\right\}}\exp{\left\{  -\frac{\gamma}{\delta}Sy +H_0^{(j)}(S,t)+\varepsilon^{\frac23}H_2^{(j)}(S,t) +O(\varepsilon) \right\}},\nonumber
\end{eqnarray}
where the absolute value of the
$O(\varepsilon)$ term is bounded by $\varepsilon$
times a constant that is independent of $\varepsilon>0$ and $(t,B,y,S)$,
but may depend on the compact $\K$.

Moreover, let a strategy $(\tilde L^{(j)}, \tilde M^{(j)}),~j=1,w$, be the strategy associated with $\NT$ region. This strategy is ``nearly optimal", i.e. for $(t, B,y,S)\in\zeroClosedTclosed\times\R\times\K$ and for $j=1,w$ the expectation of the utility of the final wealth for this strategy for $\varepsilon>0$ small enough satisfies 
\begin{eqnarray}
&&\tilde V^{(j)}(t,B,y,S)\define\E_t^{B,y,S} \left[\U ( \Phi^{(j)}(T,\tilde B_T^{(j)}, \tilde y_T^{(j)},S_T))\right]\label{eq:tilde-V-expan}\\
&&=\Util-\exp{ \left\{-\gamma \frac{B}{\delta}\right\}}\exp{\left\{  -\frac{\gamma}{\delta}Sy +H_0^{(j)}(S,t)+\varepsilon^{\frac23}H_2^{(j)}(S,t) +O(\varepsilon) \right\}},\nonumber
\end{eqnarray}
where $(s, \tilde B_s^{(j)}, \tilde y_s^{(j)},S_s),~ t\le s\le T$ is the process associated with this trading strategy. In other words, the logarithm of the expected utility of the final payoff of this strategy matches the logarithm of the value function at the order $\varepsilon^{\frac23}$.
\end{theorem}

Using utility indifference pricing, Davis,  Panas  \& Zariphopoulou \cite{DavisPanasZariphopoulou} in Theorem 1 show that the price of the contingent claim liability $V(S,t)$ is given by
\begin{equation}
V(S,t)=\frac{\delta}{\gamma}\log\left( \frac{Q^{(w)}(S,0,t)}{Q^{(1)}(S,0,t)}\right),~(S,t)\in\Rpplus\times\zeroClosedTclosed.
\label{eq:price}
\end{equation}

It follows as simple corollary of Theorem \ref{thm:Thm2} that 
\begin{Cor}\label{cor:main}
Assume $ \gamma>0$, and assume that $g$, the payoff of a European style contingent claim with maturity $T>0$, satisfies Assumptions \ref{A1} and \ref{A2}. Fix a compact $\K\subset\Rpplus$. Then for $(S,t)\in\K\times\zeroClosedTclosed$ the price of the contingent claim $V(S,t)$ is given for $\varepsilon>0$ small enough by
\begin{equation}
V(S,t)=V_0(S,t)+\frac{\delta}{\gamma}\varepsilon^{\frac23}\left( H_2^{(w)}(S,t)-H_2^{(1)}(S,t)\right)+O(\varepsilon),
\label{eq:V}
\end{equation}
where the absolute value of the
$O(\varepsilon)$ term is bounded by $\varepsilon$
times a constant that is independent of $\varepsilon>0$ and $(S,t)$,
but may depend on the compact $\K$.
\end{Cor}

In order to prove Theorem \ref{thm:Thm2}, we use the following two auxiliary theorems:

\begin{theorem}\label{Thm1}
Assume $\varepsilon, \gamma>0$, and assume that $g$, the payoff of a European style contingent claim with maturity $T>0$, satisfies Assumptions \ref{A1} and \ref{A2}. Fix a compact $\K\subset\Rpplus\times\R$. Then there exist functions $\Qjpm \in C^{2,1,1}(\Rplus\times\R\times\zeroClosedTclosed),~j=1,w$ %
with the following properties:
\begin{enumerate}
\item $\pm \mathcal H_3 \Qjpm \ge 0$ in $\Rplus\times\R\times\zeroClosedTclosed.$
\item $\pm \Qjpm(S,y,T) \le \mp\U(\Phi^{(j)}(T, 0, y, S)),~j=1,w$. 
\item $\Qjpm$ have the following asymptotic power expansion on $\K\times\zeroClosedTclosed$:
\begin{equation}
\Qjpm(S,y,t)= \exp{\left\{  -\frac{\gamma}{\delta}Sy +H_0^{(j)}(S,t)+\varepsilon^{\frac23}H_2^{(j)}(S,t) +O(\varepsilon) \right\}}.
\label{eq:Qjpm-expan}
\end{equation}
\end{enumerate}
Here, again
$O(\varepsilon)$ term is bounded by $\varepsilon$
times a constant that is independent of $\varepsilon>0$ and $(S,y,t)$,
but may depend on the compact $\K$.
\end{theorem}

Our next goal is to prove that $Q^{\jm}$ dominates $Q^{\j}$ which in turn dominates $Q^{\jp}$, $j=1,w$. 
However, first for $(t,B,y,S)\in\zeroClosedTclosed\times\R\times\R\times\Rplus$ we define 
\begin{equation}
\psi^{\jpm}(t,B,y,S) \define\Util-\exp{\left\{-\gamma\frac{B}{\delta}\right\}}Q^{\jpm}(S,y,t),~j=1,w.
\label{eq:psi}
\end{equation}

It turns out that it is easier to show that $\pm\psi^{\jpm}\ge \pm V^{\j},$ then $Q^{\jp} \le Q^{\j}\le Q^{\jm}$. The later, then follows, if we let $B=0$. This is a corollary of the second auxiliary theorem.

\begin{theorem}\label{thm:comp}
Assume the assumptions of Theorem \ref{thm:Thm2} hold. For $j=1,w$ let $\psi^{\jpm}$ be the functions defined in \eqref{eq:psi}. %
For $(t,B,y,S)\in\zeroClosedTclosed\times\R\times\R\times\Rplus$ and for an admissible policy in $\mathcal T^{(j)}(t,B,y,S)$, let $(B_s, y_s, S_s),~s\in\tClosedTclosed$ be the process defined by \eqref{eq:dB}--\eqref{eq:dS}. Then for $\varepsilon>0$ small enough $\psi^{\jp}(s, B_s, y_s, S_s),~s\in\tClosedTclosed$ is a supermartingale. 

Moreover, for the ``nearly-optimal" strategy $(\tilde L^{\j}, \tilde M^{\j}),~j=1,w$, the strategy associated with $\NT$ region, let the associated process be $(\tilde B_s^{\j}, \tilde y_s^{\j}, S_s),~s\in\tClosedTclosed$. Then $\psi^{\jm}(s, \tilde B_s^{\j}, \tilde y_s^{\j}, S_s)$ is a submartingale. This theorem is often referred to as the verification argument.
\end{theorem}

{\sc Proof of Theorem \ref{thm:Thm2}:}
Let $Q^{\jpm}$ be the functions constructed in Theorem \ref{Thm1} and $\psi^{\jpm}$ defined in \eqref{eq:psi}. 
For $j=1,w, ~(t,B,y,S)\in\zeroClosedTclosed\times\R\times\R\times\Rplus$ we have
\begin{eqnarray*}
\psi^{\jp}(t,B,y,S)&\ge& \E_t^{B,y,S}\left[-\exp { \left\{-\gamma B_T\right\} } Q^{\jp}(S_T,y_T,T)\right]\ge \E_t^{B,y,S} \left[\U ( \Phi^{(j)}(T, B_T,  y_T,S_T))\right],
\end{eqnarray*}
where we have used the fact that $\psi^{\jp}$ is a supermartingale, to establish the first inequality, and the final time condition of $Q^{\jp}$ and \eqref{eq:Phi}, \eqref{eq:Q-final-time} to conclude the second inequality. Maximizing over all admissible strategies, we conclude that
$$\psi^{\jp}(t,B,y,S) \ge V^{\j}(t,B,y,S).$$

Similarly, 
\begin{eqnarray*}
\psi^{\jm}(t,B,y,S)&\le& \E_t^{B,y,S}\left[-\exp { \left\{-\gamma \tilde B_T^{\j}\right\} } Q^{\jm}(S_T,\tilde y_T^{\j},T)\right] \le \E_t^{B,y,S} \left[\U ( \Phi^{(j)}(T, \tilde B_T^{\j},  \tilde y_T^{\j},S_T))\right].
\end{eqnarray*}

Then, 
$$\psi^{\jm}(t,B,y,S) \le \E_t^{B,y,S} \left[\U ( \Phi^{(j)}(T, \tilde B_T^{\j},  \tilde y_T^{\j},S_T))\right]\le V^{\j}(t,B,y,S).$$
The desired asymptotic power expansion \eqref{eq:V-expan} follows from \eqref{eq:Qjpm-expan}. Moreover, we conclude that \eqref{eq:tilde-V-expan} holds and that the strategy $(\tilde L^{(j)}, \tilde M^{(j)}),~j=1,w$ is ``nearly optimal" as logarithm of the expected utility of the final payoff of this strategy matches the logarithm of the value function at $O(\varepsilon^{\frac23})$.

$\hfill\Box$
\section{Discussion of the Results}
\setcounter{equation}{0} \label{sec:discussion}
In this section we analyze the case of European call option. However, first we need a few remarks.
\begin{remark}\label{remark:y-star}
From \eqref{eq:H_0}, \eqref{eq:y-star} in case without contingent claim liability, $j=1$,  
\begin{equation}
H_{0_S}^{(1)}=0,\mbox{ so }y^{(1)^{*}}=\frac{\delta(\mu-r)}{\gamma S\sigma^2},~y^{(1)^{*}}_S=-\frac{\delta(\mu-r)}{\gamma S^2\sigma^2}.
\label{eq:no-option}
\end{equation}
In case of having an contingent claim liability 
$\frac{\delta}{\gamma} H_{0_S}^{(w)}= V_{0_S}=\Delta(S,t)$, and $\frac{\delta}{\gamma} H_{0_{SS}}^{(w)}= V_{0_{SS}}=\Gamma(S,t)$ where $\Delta(S,t)$ and $\Gamma(S,t)$ are the delta and the gamma of the contingent claim in standard Black-Scholes model. In this case 
\begin{equation}y^{(w)^{*}}=\Delta(S,t)+\frac{\delta(\mu-r)}{\gamma S\sigma^2},~y^{\ws}_S(S,t) = \Gamma(S,t)-\frac{\delta(\mu-r)}{\gamma S^2\sigma^2}.
\label{eq:with-option}
\end{equation}
\end{remark}

\begin{lemma}\label{lemma:gamma-bounds}
Assume Assumption \ref{A1} holds. Then $V_0-SV_{0_S},~S^2V_{0_{SS}}, ~S^3V_{0_{SSS}}$ and $S^4V_{0_{SSSS}}$ are bounded in $\Rpplus\times\zeroClosedTclosed$ by $\sup_{ x>0} \abs{g(x)-g'(x)x}$, $ \sup_{x>0}x^2\abs{g''(x)}$,  $\sup_{x>0}\abs{x^3g^{(3)}(x)}$ and $\sup_{x>0}x^4\abs{g^{(4)}(x)}$ respectively. Moreover, $S^2y_{S}^{\jstr}, ~S^3y_{SS}^{\jstr},$ $S^4y_{SSS}^{\jstr}, ~S^2y_{St}^{\jstr},j=1,w,$ and $SV_{0_{St}},S^2V_{0_{SSt}}$ are bounded in $\Rpplus\times\zeroClosedTclosed.$ 
\end{lemma}

The proof is differed until the Appendix.

\begin{example}\label{ex:option}
The standard examples of a call or a put option payoff with strike $K$, $(S_T-K)^{+}$ and $(K-S_T)^{+}$ do not satisfy Assumption \ref{A1}. Instead for $(S,\tau)\in \Rplus\times\Rplus$ consider the Black-Scholes call option price
\begin{equation}
C_{BS}(S,\tau)=SN(d_{+}(S,\tau))-\e{-r\tau}KN(d_{-}(S,\tau)),
\label{eq:call-price}
\end{equation}
where $N(x)$ is the density of a standard normal variable and
\begin{equation}
d_{\pm}(S, \tau)=\frac{\log\left(\frac{S}{K}\right)+\left(r\pm\frac{\sigma^2}{2}\right)\tau}{\sigma\sqrt{\tau}}.
\label{eq:d_pm}
\end{equation}
For fixed $\Delta T>0$ let $g(S)\define C_{BS}(S,\Delta T)$. It is well known that $g(S)> (S-K)^{+}$ and $g(S)$ converges to $(S-K)^{+}$ as $\Delta T\rightarrow 0$ uniformly in $S$. It is not hard to verify that $g$ satisfies Assumptions \ref{A1}, since we can easily compute all of its derivatives. 
Using the fact that $g''(S)=\frac{N'(d_{+}(S, \Delta T))}{S\sigma\sqrt{\Delta T}}$, it follows $\abs{g''} \le \frac{C}{\sqrt{\Delta T}}$ in $\Rpplus$, for some constant $C>0$. Hence we can choose $\Delta T >0$ so that Assumption \ref{A2} will be satisfied, however, an optimal $\Delta T$ does not exist.
From Assumption \ref{A2} and Lemma \ref{lemma:gamma-bounds} we see that the cash gamma of the option $S^2V_{0_{SS}}$ cannot exceed $\frac{\delta(\mu-r)}{\gamma \sigma^2}$, %
which is useful in practical applications. In this example the delta hedge also has a very simple interpretation - at time $t$ one simply hedges with delta of an option with a slightly longer maturity $\Delta T +(T-t)$.

Assumption \ref{A1} has a number of requirements. The requirement that the payoff function $g$ is smooth is needed so  there will be a delta hedge, at least for the case of zero transactions costs. The second requirement that $g(S)-g'(S)S$ is bounded in $\Rpplus$, helps ensure a bound from below of the wealth of any admissible strategy in case of having a contingent claim. Finally, we also need to have bounded cash gamma and higher derivatives, in order to insure that the cash delta of the contingent claim would not have big swings in order to control the amount of transaction costs paid.

Assumption \ref{A2} is also natural. Indeed, in case of a call option, we know that close to maturity when the stock is close to the option strike, the (cash) gamma of the option is big. This means that when the stock moves the (cash) delta can make big swings. This requires the agent to either be (risk) tolerant to such big movements away from the optimal strategy, or trade more in order to hedge the option closely and incur big transaction costs. This is exactly what Assumption \ref{A2} says. Assuming all the other model parameters are fixed, either the maximum (cash) gamma of the option is small, or the agent's risk aversion coefficient $\gamma$ is small.

Another reason why Assumption \ref{A2} is critical, i.e. $\Delta T$ needs to be strictly positive, is that it can be shown that as $\Delta T \searrow 0$ the $O(\varepsilon)$ term in \eqref{eq:V} diverges. See Remark \ref{remark:H4-explosion} for details.
Hence we need to make a mollification of the sort $g(S)= C_{BS}(S,\Delta T)$ to the true call payoff $(S-K)^{+}$ in order for the power expansion \eqref{eq:V-expan} to be valid.
Interestingly, the limit $\lim\limits_{\Delta T \searrow 0}H_2^{(w)}$ is well defined.

\begin{figure}[htbp]
\begin{center}
\caption{Plot of $\H_2$ as a function of $\Delta T$, with model parameters and initial values $S=1,K=1, T=1, r=0,\mu=0.1,\sigma=\sqrt{2},
\gamma=1.$}
\epsfysize=0.5\textwidth\epsffile{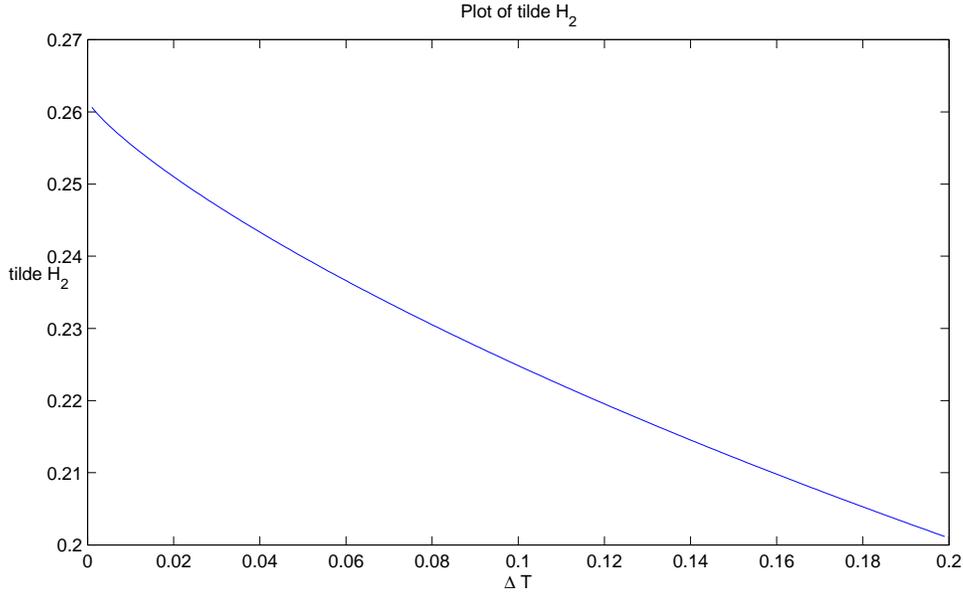}
\label{fig:H_2}
\end{center}
\end{figure}
Indeed, for $(S,t)\in\Rplus\times\zeroClosedTclosed$, let $f(S,t)\define\frac12\left(\frac{3\gamma^2S^4\sigma^3{y^{\ws}_S}^{2}}{2\delta^2} \right)^{\frac23}$, it is shown in Lemma \ref{lemma:bounds} that
$$H_2^{\w}(S,t)  =\e{-\frac12(k-1)x-\frac14(k+1)^2\tau+r(T-t)}\H_2(\tau,x),$$ 
where $k=\frac{2r}{\sigma^2},~S=\e{x},~\tau=\frac{\sigma^2}{2}(T-t)$, and where
\begin{equation}
\H_2(\tau,x)=\frac2{\sigma^2}\int_0^\tau\int_{-\infty}^{\infty}\Psi(x-y,\tau-s)\e{\frac12(k-1)y+\frac14(k+1)^2s-ks}f(\e{y},T-\frac2{\sigma^2}s)dyds,
\label{eq:heat-eq-sol1-tmp1}
\end{equation}
with $\Psi(x,t)=\frac1{\sqrt{4\pi t}}\exp{\left\{-\frac{x^2}{4t}\right\}}$ is the heat kernel.

We show that $\lim\limits_{\Delta T \searrow 0}\H_2^{(w)}$ exists. From \eqref{eq:with-option} in Remark \ref{remark:y-star} we calculate that $y^{\ws}_S(S,t) = \frac{N'(d_{+}(S,\Delta T+T-t))}{\sigma S\sqrt{\Delta T +T-t}}-\frac{\delta(\mu-r)}{\gamma S^2\sigma^2}$. It follows that %
$\abs{f(S,t)}\le C\left(1+\left(T-t\right)^{-\frac23}\right)$ for some constant $C>0$ and uniformly in $S\in\Rpplus.$ %
Clearly $\H_2(0,x)=0.$ Moreover, it follows from \eqref{eq:heat-eq-sol1-tmp1} that for $(\tau,x)\in (0,\frac{\sigma^2}2T]\times\R$ and $0<\varepsilon_2<1\wedge\tau$, we have
\begin{eqnarray*}
\abs{\H_2\left(\tau,x\right)}&\le&\frac{2}{\sigma^2}\int_0^\tau\int_{-\infty}^{\infty}\Psi(x-y,\tau-s)\e{\frac12(k-1)y}\abs{f(\e{y},T-\frac2{\sigma^2}s)}dyds\\
&\le&C\int_0^\tau\int_{-\infty}^{\infty}\Psi(x-y,\tau-s)\e{\frac12(k-1)y}    \left(1+\left(\frac2{\sigma^2}s\right)^{-\frac23}\right) dyds\\
&\le& C\int_0^\tau\int_{-\infty}^{\infty}\Psi(x-y,\tau-s)\e{\frac12(k-1)y}     dyds\\
&&\quad+ C\int_{\varepsilon_2}^\tau\int_{-\infty}^{\infty}\Psi(x-y,\tau-s)\e{\frac12(k-1)y} dyds\\
&&\quad+C\int_0^{\varepsilon_2}\left(\frac2{\sigma^2}s\right)^{-\frac23}ds~ {\varepsilon_2}\int_{-\infty}^{\infty}\Psi(x-y,\tau)\e{\frac12(k-1)y}     dy
\end{eqnarray*}
is finite. In the above inequalities, $C$ denotes a generic constant, that may differ from line to line. We conclude that $H_2^{\w}(S,t)$ is uniformly bounded in $\Delta T\in\Rpplus$. 

As seen in Figure \ref{fig:H_2}, $\H_2$ (and thus also $H_2^{(w)}$) increases as $\Delta T$ decreases to zero. Clearly, as $\Delta T$ decreases to zero, the gamma of the call option increases around $S_T=K$, which in turn increases the hedging transaction costs. Even though the price of the contingent claim liability without transaction costs decreases, since the payoff $g$ decreases with $\Delta T$, the overall price with transaction costs increases as $\Delta T$ goes to zero (at least at the $O(\varepsilon^\frac23$)).

\end{example}

\section{Rigorous Power Expansion}
\setcounter{equation}{0} \label{sec:expansion}
In the following section we present a few more definitions and preliminary results that would lead to the proof of Theorem \ref{Thm1}. Specifically, we would define the functions $Q^{\jpm}$, show their smoothness, and establish important bounds of the terms used in the construction of $Q^{\jpm}$.

The reader is reminded of the definitions of $y^{\jstr}, Y^{\j}, \NT$ defined in \eqref{eq:y-star}, \eqref{eq:y_Ypm}, \eqref{eq:NT}.
Similar to Whalley \& Wilmott \cite{WhalleyWilmott}, for fixed $j\in\{1,w\}$, we will also use the $y$ coordinate translation
\begin{equation}
y=\yjstr+\varepsilon^{\frac13}Y.
\label{eq:y2Y}
\end{equation}
One can also think of $Y$ as a function of $(S,y,t)\in\overline\NT$ given by 
\begin{equation}
Y(S,y,t)\define (y-\yjstr)\varepsilon^{-\frac13}.
\label{eq:y2Y-func}
\end{equation}
For convenience we will often drop the arguments and refer to $Y(S,y,t)$ simply as $Y$.
 
Thus for a smooth function $h:\overline\NT\rightarrow \R$ for $j=1,w$ we will have the chain rule:
\begin{equation}
\left\{
\begin{array}{ccl}
\frac{\partial h}{\partial y}(S,y,t)&=&\varepsilon^{-\frac13}\frac{\partial h}{\partial Y}(S,Y,t),\\
\frac{\partial h}{\partial S}(S,y,t)&=&\frac{\partial h}{\partial S}(S,Y,t)-\varepsilon^{-\frac13}y_S^{\jstr}\frac{\partial h}{\partial Y}(S,Y,t),\\
\frac{\partial h}{\partial t}(S,y,t)&=&\frac{\partial h}{\partial t}(S,Y,t)-\varepsilon^{-\frac13}y_t^{\jstr}\frac{\partial h}{\partial Y}(S,Y,t).
\end{array}
\right.
\label{eq:chain-rule}
\end{equation}
Inspired by the coordinate translation, the {\em``no-trade"} region for $j=1,w$, in the translated $Y$ coordinate, would be defined as 
\begin{equation}
\NTY\define\Big\{(S,Y,t)\in\Rplus\times\R\times\zeroClosedTclosed \colon \abs{Y} < Y^{\j}\Big\}.
\label{eq:NT_Y}
\end{equation}
For convenience, for $j=1,w$, we also define the buy and the sell regions as
\begin{eqnarray}
\B&\define&\Big\{(S,y,t)\in\Rplus\times\R\times\zeroClosedTclosed \colon y< y^{\jm}\Big\},\label{eq:buy}\\
\S&\define&\Big\{(S,y,t)\in\Rplus\times\R\times\zeroClosedTclosed \colon y> y^{\jp}\Big\}.\label{eq:sell}
\end{eqnarray}
\begin{remark}\label{remark:NT-width}
For $(S,t)\in \Rplus\times\zeroClosedTclosed$ we have that $SY^{\j}=\left(\frac{3S^4\delta (y_S^{\jstr})^2}{2\gamma} \right)^{\frac13}$, for $j=1,w$. From Lemma \ref{lemma:gamma-bounds} $S^2y_s^{\jstr}$ is uniformly bounded in $(S,t)$ in either case with and without contingent claim liability, thus so is $SY^{\j}$. Additionally, we conclude that the width, in $Y$, of the $\overline\NTY$ region at $(S,t)$ equal to $2Y^{\j}$, when scaled by $S$ is also uniformly bounded in $(S,t)$ in both cases $j=1,w$. Moreover, for $j=1$, $S^4\left(y^{\ones}_S\right)^2=\frac{\delta^2(\mu-r)^2}{\gamma^2\sigma^4}$ 
is strictly positive, the same is true for $j=w$, since from \eqref{eq:with-option} 
$S^4\left(y^{\ws}_S\right)^2=\left(\frac{\delta(\mu-r)}{\gamma\sigma^2}-S^2V_{0_{SS}}\right)^2\ge\left(\frac{\delta(\mu-r)}{\gamma\sigma^2}-\sup\limits_{x\in[0,\infty)}x^2g''(x) \right)^2\ge  \varepsilon_1^2$,
where we have also used Assumption \ref{A2} and that
$\abs{S^2V_{0_{SS}}} \le  \sup_{x>0}x^2\abs{g''(x)}$ 
from Lemma \ref{lemma:gamma-bounds}. We conclude that the width in $Y$ of the $\overline\NTY$ at $(S,t)$  when scaled by $S$ is bounded and is also bounded away from zero, in both cases, $j=1,w$. 
Similarly, the $\overline\NT$ region is centered at $\yjstr$, and its width in $y$ at $(S,t)$ equal to $y^{\jp}-y^{\jm}$, when scaled by $S$ is bounded. Its width is also bounded away from zero, and is of order $\varepsilon^{\frac13}$. Moreover, it follows that $\yjstr_S\ne 0,~j=1,w$ on $\Rpplus\times\zeroClosedTclosed$.
\end{remark}

For $(S,t)\in\Rplus\times\zeroClosedTclosed$ and $j=1,w$ we also define the function
\begin{equation}
H_3^{\jpm}(S,t)\define\mp M^{\j}(T-t)\mp M^{\j}_1,
\label{eq:H_3-def}
\end{equation}
where $M^{\j}$ is a positive constant, and
\begin{equation}
M^{\j}_1\define 4\frac{\e{ rT } (\mu-r)}{\sigma^2} +2%
, ~j=1,w.
\label{eq:M_1}
\end{equation}
In addition for $(S,Y,t)\in\overline\NTY$ and $j=1,w$ we let
\begin{equation}
H_4^{\j}(S,Y,t)%
=\frac{Y^2}{2}\left(\frac{3\gamma^2S}{2\delta^2y_S^{\jstr}} \right)^{\frac23}-\frac{\gamma^2Y^4}{12\delta^2\left({y_S^{\jstr}}\right)^{2}}.\label{eq:H_4}
\end{equation}

Define $Q^{\jpm}$ on $\overline\NT$ by
\begin{eqnarray}
&&\!\!\!\!\!\!\!\!\!\!\!\!\!\!Q^{\jpm}(S,y,t)\define \operatorname{exp} \left\{-\frac{\gamma S y }{\delta}+H_0^{\j}(S,t)+ \varepsilon^{\frac23} H_2^{\j}(S,t)\right.\label{eq:Q-pm1}\\
&& \quad\left.+\varepsilon H_3^{\jpm}(S,t)+\varepsilon^{\frac43}H_4^{\j}(S,Y,t)\right\},~j=1,w,\nonumber
\end{eqnarray}
where we have used \eqref{eq:y2Y-func} $Y=(y-\yjstr)\varepsilon^{-\frac13}$. 
We extend $Q^{\jpm}$ outside the $\NT$ region as 
\begin{eqnarray}
Q^{\jpm}(S,y,t)&=&\operatorname{exp}\left\{-\frac{\gamma(1+\varepsilon)S}{\delta}\left(y-y^{\jm}\right)\right\}Q^{\jpm}(S,y^{\jm},t),\nonumber\\
&&\qquad\qquad (S,y,t)\in\B, \label{eq:Q_buy}\\
Q^{\jpm}(S,y,t)&=&\operatorname{exp}\left\{-\frac{\gamma(1-\varepsilon)S}{\delta}\left(y-y^{\jp}\right)\right\}Q^{\jpm}(S,y^{\jp},t),\nonumber\\
&&\qquad\qquad (S,y,t)\in\S.~~~~~~~~ \label{eq:Q_sell}
\end{eqnarray}

\begin{remark}\label{remark:HJB-zero}
Note that this extension is done so that $Q^{\jpm}_y+\frac{(1+\varepsilon)\gamma SQ^{\jpm}}{\delta}=0$ in $\B,~j=1,w$, and thus, once we show $Q^{\jpm}$ is smooth, we can conclude that \eqref{eq:HJB-WW} $\mathcal H_3 Q^{\jpm}\le0$ there. Similar result holds for $\S$.
\end{remark}
\begin{remark}\label{remark:H4-explosion}
We note here that if $S^2\Gamma$ in \eqref{eq:with-option}, %
the cash gamma of the contingent claim, is unbounded on $\Rpplus\times\zeroClosedTopen$, then so are $S^2y^{\ws}_S$ and $SY^{\w}$ on $\Rpplus\times\zeroClosedTopen$. This results in $H_4^{\w}$ being unbounded on $\overline\NTYw$ as well. The unboundedness of the $O(\varepsilon)$ term in \eqref{eq:V} follows from the unboundedness of the $O(\varepsilon)$ term in \eqref{eq:V-expan} which in turn follows from the unboundedness of the $O(\varepsilon)$ term in \eqref{eq:Qjpm-expan} by similar construction as in Theorem \ref{thm:Thm2}.
\end{remark}

Differentiating \eqref{eq:H_4} we find for $j=1,w$ that in $\overline\NTY$
\begin{eqnarray}
H_{4_{Y}}^{(j)}&=&Y\left(\frac{3\gamma^2S}{2\delta^2y_S^{\jstr}} \right)^{\frac23} -
\frac{\gamma^2Y^3}{3\delta^2(y_S^{\jstr})^2},\label{eq:H_4_Y}\\
H_{4_{YY}}^{(j)}&=&\left(\frac{3\gamma^2S}{2\delta^2y_S^{\jstr}} \right)^{\frac23} -
\frac{\gamma^2Y^2}{\delta^2(y_S^{\jstr})^2}.
\label{eq:H_4_YY}
\end{eqnarray}

Furthermore, differentiating \eqref{eq:H_4} we find that in $\overline\NTY$
\begin{eqnarray}
\!\!\!\!\!\!\!\!\!\!\!\!\!\!\!\!\!\!H_{4_{S}}^{(j)}&=&\frac{Y^2}{3S^{\frac13}}\left(\frac{3\gamma^2}{2\delta^2y_S^{\jstr}} \right)^{\frac23} -\frac{Y^2}{3}\left(\frac{3\gamma^2S}{2\delta^2} \right)^{\frac23}  \left(y_S^{\jstr}\right)^{-\frac53}y_{SS}^{\jstr}+
\frac{\gamma^2Y^4}{6\delta^2(y_S^{\jstr})^3}y_{SS}^{\jstr},\label{eq:H_4_S}\\
\!\!\!\!\!\!\!\!\!\!\!\!\!\!\!\!\!\!H_{4_{YS}}^{(j)}&=&\frac{2Y}{3S^{\frac13}}\left(\frac{3\gamma^2}{2\delta^2y_S^{\jstr}} \right)^{\frac23} -\frac{2Y}{3}\left(\frac{3\gamma^2S}{2\delta^2} \right)^{\frac23}  \left(y_S^{\jstr}\right)^{-\frac53}y_{SS}^{\jstr}+
\frac{2\gamma^2Y^3}{3\delta^2(y_S^{\jstr})^3}y_{SS}^{\jstr},\label{eq:H_4_YS}\\
\!\!\!\!\!\!\!\!\!\!\!\!\!\!\!\!\!\!H_{4_{SS}}^{(j)}&=&-\frac{Y^2}{9S^{\frac43}}\left(\frac{3\gamma^2}{2\delta^2y_S^{\jstr}} \right)^{\frac23} 
-\frac{4Y^2}{9S^{\frac13}}\left(\frac{3\gamma^2}{2\delta^2} \right)^{\frac23}  \left(y_S^{\jstr}\right)^{-\frac53}y_{SS}^{\jstr}\label{eq:H_4_SS}\\
&&\quad+\frac{Y^2}{3}\left(\frac{3\gamma^2S}{2\delta^2} \right)^{\frac23} \left(y_S^{\jstr}\right)^{-\frac83}\left[ \frac53\left(y_{SS}^{\jstr}\right)^2 -y_S^{\jstr}y_{SSS}^{\jstr} \right]\nonumber%
+\frac{\gamma^2Y^4}{6\delta^2\left(y_S^{\jstr}\right)^3}\left[-\frac{3\left(y_{SS}^{\jstr}\right)^2}{y_S^{\jstr}}+y_{SSS}^{\jstr}\right],\nonumber\\
\!\!\!\!\!\!\!\!\!\!\!\!\!\!\!\!\!\!H_{4_t}^{(j)}&=&-\frac{Y^2}{2}\left(\frac{3\gamma^2S}{2\delta^2y_S^{\jstr}} \right)^{\frac23}\left(\frac43r+\frac{2y_{St}^{\jstr}}{3y_S^{\jstr}} \right)
+\frac{\gamma^2Y^4}{12\delta^2\left(y_S^{\jstr}\right)^{2}}\left(2r+\frac{2y_{St}^{\jstr}}{y_S^{\jstr}} \right).\label{eq:H_4_t}
\end{eqnarray}

\begin{remark}\label{remark:H_4_Y}
For $j=1,w$ and for $(S,t)\in\Rpplus\times\zeroClosedTclosed$ from \eqref{eq:H_4_Y}, \eqref{eq:H_4_YY} and \eqref{eq:H_4_YS} it follows that 
\begin{eqnarray}
H\ju_{4_Y}(S,\pm Y^{\j},t)&=&\pm\frac{\gamma S}{\delta},\label{eq:H_4_Y-pm}\\
H\ju_{4_{YY}}(S,\pm Y^{\j},t)&=&0,\label{eq:H_4_YY-pm}\\
H\ju_{4_{YS}}(S,\pm Y^{\j},t)&=&\pm\frac{\gamma}{\delta}.\label{eq:H_4_YS-pm}
\end{eqnarray}
Moreover, we have that $H\ju_{4_{YY}}\ge0$ in $\overline\NTY$. It follows that $\abs{H\ju_{4_Y}(S,Y,t)}\le\frac{\gamma S}{\delta}$ for $(S,Y,t)\in\overline\NTY$. 
\end{remark}

\begin{lemma}\label{lemma:bounds}
For $j=1,w$, $H_2^{\j}$, the solution to \eqref{eq:H_2} on $\Rplus\times\zeroClosedTclosed$ is given by
\begin{eqnarray}
H_2^{(1)}(S,t)&=&\left(\frac{3}{2\sigma}\right)^{\frac23}  \frac{(\mu-r)^{\frac43}(T-t)} {2},
\label{eq:H_2-sol}\\
H_2^{\w}(S,t)  &=&\e{-\frac12(k-1)x-\frac14(k+1)^2\tau+k\tau}\H_2(\tau,x),\label{eq:H_2_w-sol}
\end{eqnarray}
with
$k=\frac{2r}{\sigma^2},~S=\e{x},~\tau=\frac{\sigma^2}{2}(T-t)$, and where
\begin{equation*}
\H_2(\tau,x)=\frac2{\sigma^2}\int_0^\tau\int_{-\infty}^{\infty}\Psi(x-y,\tau-s)\e{\frac12(k-1)y+\frac14(k+1)^2s-ks}f(\e{y},T-\frac2{\sigma^2}s)dyds,
\end{equation*}
with $\Psi(x,t)=\frac1{\sqrt{4\pi t}}\exp{\left\{-\frac{x^2}{4t}\right\}}$ is the heat kernel and $f(S,t)\define\frac12\left(\frac{3\gamma^2S^4\sigma^3{y^{\ws}_S}^{2}}{2\delta^2} \right)^{\frac23}$ defined on $\R\times\Rpplus$ and $\Rplus\times\zeroClosedTclosed$ respectively.

Moreover, $H_2^{(j)}$ is bounded on $\Rpplus\times\zeroClosedTclosed$ and so are $H_4^{(j)},H_{4_{YS}}^{\j},H_{4_{t}}^{\j} $ on $\overline\NTY$. Additionally the terms $H_{2_S}^{(j)},~H_{4_S}^{(j)}$ and $y_S^{\jstr}H_{4_Y}^{\j}$, if scaled by $S$  and the terms $H_{4_{SS}}^{(j)},$ $y_S^{\jstr}H_{4_{YS}}^{\j}, \left(y_S^{\jstr}\right)^2H_{4_{YY}}^{\j}, y_{SS}^{\jstr}H_{4_{Y}}^{\j}$ if scaled by $S^2$ are all bounded in the closure of their respective domains. Moreover, for $j=1,w$ there exists $M^{\j}$ a positive constant, such that the function $H_3^{\jpm}$ defined in \eqref{eq:H_3-def} satisfies on $\Rpplus\times\zeroClosedTclosed$.
\begin{eqnarray}
&&\!\!\!\!\!\!\!\!\!\!\!\!\!\pm\left(H_{3_t}^{\jpm}+rSH_{3_S}^{\jpm}+\frac{\sigma^2S^2}{2}H_{3_{SS}}^{\jpm}\right)\ge%
\pm\max\limits_{\abs{Y}\le Y^{\j}} 
\abs{\frac{\sigma^2S^2\gamma Y}{\delta}H_{2_S}^{(j)}-\sigma^2Sy_S^{\jstr}H_{4_Y}^{(j)}+\sigma^2S^2y_S^{\jstr}H_{4_{YS}}^{(j)}} \pm1,~j=1,w.~~\label{eq:H_3}
\end{eqnarray}
Additionally, $H_{3_S}^{\jp}\equiv0$, thus $SH_{3_S}^{\jp}, S^2H_{3_{SS}}^{\jp}$ are bounded on $\Rplus\times\zeroClosedTclosed,~j=1,w.$
\end{lemma}

The proof can be found in the Appendix.

\begin{lemma}
For $j=1,w$ the functions $Q^{\jpm}$ are $C^1(\Rplus\times\R\times\zeroClosedTclosed)$.
\label{lemma:Q_C1}
\end{lemma}

{\sc Proof:}
Fix $j$. We only need to prove that $Q^{\jpm}$ continuously differentiable across the boundaries of the $\overline\NT$ region, $j=1,w$. Consider the buy boundary first and fix $(\bar S,\bar t)\in \Rpplus\times\zeroClosedTclosed$. Then $(\bar S,y^{{\jm}},\bar t)\in \partial \B$. Differentiating \eqref{eq:Q_buy} and \eqref{eq:Q-pm1} we have
\begin{eqnarray}
Q_y^{\jpm}(S,y,t)&=& -\frac{\gamma(1+\varepsilon)S}{\delta}Q^{\jpm}(S,y,t), ~~~(S,y,t)\in \B,
\label{eq:Q_y-in-B}\\
Q_y^{\jpm}(S,y,t)&=&\left(-\frac{\gamma S}{\delta}+\varepsilon H_{4_Y}^{\j}(S,Y,t)\right)Q^{\jpm}(S,y,t), ~~~(S,y,t)\in \NT.~~~~~~~~~\label{eq:Q_y-in-NT}
\end{eqnarray}
Taking the limit in \eqref{eq:Q_y-in-B} and \eqref{eq:Q_y-in-NT} as $(S,y,t)$ approaches $(\bar S,y^{{\jm}},\bar t)$ from inside $\B$ and $\NT$ respectively, we see that these limits are equal. Indeed from \eqref{eq:y2Y} $y=y^{{\jm}}$ corresponds to $Y=-Y^{\j}$, and by Remark \ref{remark:H_4_Y} $H_{4_Y}(\bar S,-Y^{\j},\bar t)=-\frac{\gamma \bar S}{\delta}$. This shows that $Q^{\jpm}$ is continuously differentiable with respect to $y$ across the boundary of the buy region, and we conclude that
\begin{equation}
\frac{\gamma(1+\varepsilon)\bar S}{\delta}Q^{\jpm}(\bar S,y^{{\jm}},\bar t)=-Q_y^{\jpm}(\bar S,y^{{\jm}},\bar t).
\label{eq:equality-Q_y}
\end{equation}
Moreover, inside $\B$ we compute
\begin{eqnarray}
Q_S^{\jpm}(S,y,t)&=&\left(-\frac{\gamma(1+\varepsilon)}{\delta}(y-y^{\jm})+\frac{\gamma(1+\varepsilon )S}{\delta}y_S^{\jm}\right)Q^{\jpm}(S,y,t)\label{eq:Q_S-in-B}\\
&&\!\!\!\!\!\!\!\!\!\!\!\!\!\!\!+\operatorname{exp} \left\{-\frac{\gamma(1+\varepsilon)S}{\delta}(y-y^{\jm}) \right\}\left(Q_S^{\jpm}(S,y^{\jm},t) +Q_y^{\jpm}(S,y^{\jm},t)y_S^{\jm} \right),\nonumber\\
Q_t^{\pm}(S,y,t)&=&\left(r\frac{\gamma(1+\varepsilon)S}{\delta}(y-y^{\jm})+\frac{\gamma(1+\varepsilon )S}{\delta}y_t^{\jm}\right)Q^{\jpm}(S,y,t)\label{eq:Q_t-in-B}\\
&&\!\!\!\!\!\!\!\!\!\!\!\!\!\!\!+\operatorname{exp} \left\{-\frac{\gamma(1+\varepsilon)S}{\delta}(y-y^{\jm}) \right\}\left(Q_t^{\jpm}(S,y^{\jm},t) +Q_y^{\jpm}(S,y^{\jm},t)y_t^{\jm} \right),\nonumber
\end{eqnarray}
where all the derivatives of $Q^{\jpm}$ on the right hand side above should be understood as a limit of the appropriate derivative from inside $\NT$. Taking the limit of as $(S,y,t)\rightarrow (\bar S,y^{{\jm}},\bar t)$ from inside $\B$ and using \eqref{eq:equality-Q_y} we conclude that \eqref{eq:Q_S-in-B} and \eqref{eq:Q_t-in-B} converge to $Q_S^{\jpm}(\bar S,y^{{\jm}},\bar t)$ and $Q_t^{\jpm}(\bar S,y^{{\jm}},\bar t)$ respectively.

Similar calculation hold for the sell boundary $\partial \S$.

$\hfill\Box$

\begin{lemma}\label{lemma:Q_C2}
For $j=1,w$, the functions $Q^{\jpm}$ are twice continuously differentiable in $S$ on $\Rpplus\times\R\times\zeroClosedTclosed$.
\end{lemma}

{\sc Proof:}
Fix $j$. Again, we only need to prove the claim across the boundaries of the $\overline\NT$ region. Consider the buy boundary first,  fix $(\bar S, \bar t)\in\Rpplus\times\zeroClosedTclosed$. Then $(\bar S,y^{{\jm}},\bar t)\in \partial \B$. We first show that the following holds
\begin{eqnarray}
\!\!\!\!\!\!\!\!\!\!\!\!\!\! &&Q_{yS}^{\jpm}(\bar S,y^{{\jm}}, \bar t)+\frac{\gamma(1+\varepsilon)}{\delta}\left( Q^{\jpm}(\bar S,y^{{\jm}}, \bar t)+\bar SQ_{S}^{\jpm}(\bar S,y^{{\jm}}, \bar t)\right)=0,\label{eq:equality-Q_yS}\\
\!\!\!\!\!\!\!\!\!\!\!\!\!\! &&Q_{yy}^{\jpm}(\bar S,y^{{\jm}},\bar t)+\frac{\gamma(1+\varepsilon)\bar S}{\delta}\left(2Q_{y}^{\jpm}(\bar S,y^{{\jm}},\bar t)+\frac{\gamma(1+\varepsilon)\bar S}{\delta}Q^{\jpm}(\bar S,y^{{\jm}},\bar t)\right)=0,\nonumber\\
\label{eq:equality-Q_yy}
\end{eqnarray}
where all the second derivatives of $Q^{\jpm}$ above should be understood as a limit of the appropriate second derivative from inside $\NT$. Differentiating \eqref{eq:Q_y-in-NT} we calculate that for $(S,y,t)\in \NT$
\begin{equation}
Q_{yy}^{\jpm}(S,y,t)=\left[\varepsilon^{\frac23} H_{4_{YY}}\ju(S,Y,t)+\left(-\frac{\gamma S}{\delta}+\varepsilon H_{4_Y}\ju(S,Y,t)\right)^2\right]Q^{\jpm}(S,y,t)\label{eq:Q_yy-in-NT}.
\end{equation}
The limit $y\to y^{{\jm}}$ corresponds by \eqref{eq:y2Y} to $Y\to-Y^{\j}$. Using \eqref{eq:H_4_Y-pm} and \eqref{eq:H_4_YY-pm}, we see that \eqref{eq:equality-Q_yy} follows from \eqref{eq:equality-Q_y}.
Similarly for $(S,y,t)\in \NT$ differentiating \eqref{eq:Q_y-in-NT} we have
\begin{eqnarray}
&&\!\!\!\!\!\!\!\!\!\!Q_{yS}^{\jpm}(S,y,t)=\left[-\frac{\gamma}{\delta} + \varepsilon H_{4_{YS}}\ju(S,Y,t)-\varepsilon^{\frac23}H_{4_{YY}}\ju(S,Y,t)\yjstr_S\right]Q^{\jpm}(S,y,t)\label{eq:Q_yS-in-NT}\\
&&\!\!\!\!\!\!\!\!\!\!\qquad+\left[ -\frac{\gamma S}{\delta} + \varepsilon H_{4_{Y}}\ju(S,Y,t)\right]Q_S^{\jpm}(S,y,t).\nonumber
\end{eqnarray}
Again, letting $y\to y^{{\jm}}$, and using \eqref{eq:H_4_Y-pm}, \eqref{eq:H_4_YY-pm} and \eqref{eq:H_4_YS-pm} equation \eqref{eq:equality-Q_yS} follows.
We calculate $Q_{SS}^{\jpm} $ for  $(S,y,t)\in\B$, by differentiating \eqref{eq:Q_S-in-B} with respect to $S$:
\begin{eqnarray}
Q_{SS}^{\jpm}(S,y,t)&=&\left(\frac{2\gamma(1+\varepsilon)}{\delta}y_S^{\jm}+\frac{\gamma(1+\varepsilon )S}{\delta}y_{SS}^{\jm}\right)Q^{\jpm}(S,y,t)\label{eq:Q_SS-in-B}\\
&+&2\operatorname{exp} \left\{-\frac{\gamma(1+\varepsilon)S}{\delta}(y-y^{\jm}) \right\}\left(Q_S^{\jpm}(S,y^{\jm},t) +Q_y^{\jpm}(S,y^{\jm},t)y_S^{\jm} \right)\nonumber\\
&&\qquad\times
\left(-\frac{\gamma(1+\varepsilon)}{\delta}(y-y^{\jm})+\frac{\gamma(1+\varepsilon )S}{\delta}y_{S}^{\jm} \right)\nonumber\\
&+&\operatorname{exp} \left\{-\frac{\gamma(1+\varepsilon)S}{\delta}(y-y^{\jm}) \right\}
\left(-\frac{\gamma(1+\varepsilon)}{\delta}(y-y^{\jm})+\frac{\gamma(1+\varepsilon )S}{\delta}y_{S}^{\jm} \right)^2
Q^{\jpm}(S,y^{\jm},t)\nonumber\\
&+&\operatorname{exp} \left\{-\frac{\gamma(1+\varepsilon)S}{\delta}(y-y^{\jm}) \right\}\nonumber\\
&&\qquad\times
\left(Q_{SS}^{\jpm}(S,y^{\jm},t)+2Q_{Sy}^{\jpm}(S,y^{\jm},t)y_S^{\jm}+Q_{yy}^{\jpm}(S,y^{\jm},t)\left(y_S^{\jm}\right)^2%
+Q_{y}^{\jpm}(S,y^{\jm},t)y_{SS}^{\jm} \right)\nonumber.
\end{eqnarray}
Taking the limit of \eqref{eq:Q_SS-in-B} as $(S,y,t)\rightarrow (\bar S,y^{{\jm}},\bar t)$ from inside the $\B$ region and using \eqref{eq:equality-Q_y}, \eqref{eq:equality-Q_yS} and \eqref{eq:equality-Q_yy} we see that $Q_{SS}^{\jpm}(S,y,t)\rightarrow Q_{SS}^{\jpm}(\bar S,y^{{\jm}},\bar t)$. 
Similar calculation can be done for $\S$ region.
$\hfill\Box$

\section{ \emph{{Proof of Theorem \ref{Thm1}}}} \label{proofMainThm}

We are now ready to tackle the proof of Theorem \ref{Thm1}. We have already constructed the functions $Q^{\jpm}$ and shown their smoothness in Lemmas \ref{lemma:Q_C1} and \ref{lemma:Q_C2}. The rest of the proof is divided into multiple steps. We first deal with the final time condition and prove that $\pm Q^{{\jpm}}(S,y,T) \le\pm Q^{\j}(S,y,T)$ for $(S,y)\in\Rpplus\times\R$. Next we show that $\pm \mathcal H_3 \Qjpm \ge 0$ in $\Rplus\times\R\times\zeroClosedTclosed.$ The claim for $Q^{\jm}$ follows from showing that $\D Q^{\jm} \le 0$, since by Remark \ref{remark:HJB-zero} $\mathcal H_3 Q^{\jm}\le0$ on $\B$ and $\S$. We verify that $\D Q^{\jpm} \le 0$ in Step 2. We have to work harder to prove the claim for $Q^{\jp}$ and hence the asymmetry and the extra steps.

\subsection{ \emph{{Step 1: Final Time Conditions}}} \label{step1}

Let $S>0,~y\in \R$. From \eqref{eq:Q-final-time}, the final time condition is in case of no contingent claim liability with $j=1$
\begin{equation}
Q\one(S,y,T)=\exp{\{-\gamma c(y,S)\}},
\label{eq:final_cond1-WW}
\end{equation}
and in case of contingent claim liability with $j=w$ 
\begin{equation}
Q\w(S,y,T)=\exp{\{-\gamma (c(y-g'(S),S)-(g(S)-g'(S)S)) \}}.
\label{eq:final_cond2-WW}
\end{equation}
We want to set the final time conditions for $Q^{\jpm},~j=1,w$, such that 
\begin{equation}
Q^{\jp}(S,y,T)\le Q\ju(S,y,T)\le Q^{\jm}(S,y,T), ~(S,y)\in\Rpplus\times\R, ~j=1,w.
\label{eq:final-time-cond}
\end{equation}
In case $j=1$, from Remark \ref{remark:y-star} we have that $y^{(1)^{*}}(S,T)=\frac{\mu-r}{\gamma S\sigma^2}$, so $Sy^{(1)^{*}}$ is bounded. Moreover, from Remark \ref{remark:NT-width} the width of $\overline\NTo$ region in $y$ scaled by $S$ is bounded and of order $\varepsilon^{\frac13}.$ Thus for $(S,y,T)\in \overline \NTo$ we have that 
\begin{eqnarray}
-\gamma c(y,S)&=&-\gamma(1-\varepsilon\sign{y})yS=-\gamma yS+\gamma\varepsilon\sign{y}(y-y^{\ones}+y^{\ones})S~~~~~~~~\label{eq:Q-final-cond}\\
&=&-\gamma yS+\sign{y} \frac{\mu-r}{\sigma^2}\varepsilon+O(\varepsilon^{\frac43}).\nonumber
\end{eqnarray}
In this case, for $(S,y,T)\in \overline \NTo$ using the final time conditions $H_1\one(S,T)=H_2\one(S,T)=0$ and $H_3{\onepm}(S,T) = \mp M_1\one$ and the boundedness of $H_4\one$ shown in Lemma \ref{lemma:bounds}, we have that
\begin{equation}
Q^{\onepm}(S,y,T)=\exp{\left\{-\gamma yS  \mp \varepsilon M_1\one+O(\varepsilon^{\frac43})   \right\}}.
\label{eq:Q-final-cond1}
\end{equation}
Comparing \eqref{eq:Q-final-cond1} with exponentiated \eqref{eq:Q-final-cond}, we conclude from the definition of $M^{\one}_1$ that for $\varepsilon>0$ small enough %
$\pm Q^{(1)^{\pm}}(S,y,T) \le\pm \exp{ \left\{ \mp\frac\epsilon2 M_1^{\one}\right\}  } Q^{\one}(S,y,T)\le \pm Q^{\one}(S,y,T).$

In case of contingent claim liability, $j=w$, it follows from Remark \ref{remark:y-star} that $ y^{\w}(S,T) = g'(S)+\frac{\mu-r}{\gamma\sigma^2S}.$ For $(S,y,T)\in \overline \NTw$ it follows from Remark \ref{remark:NT-width} that $\left(y-y\ws\right)S$ is bounded and of order $\varepsilon^{\frac13}$.
We calculate that
\begin{eqnarray}
&&-\gamma\left[c(y-g'(S),S)-(g(S)-g'(S))S\right]\nonumber\\
&&=-\gamma\left[yS-g'(S)S-\varepsilon \sign{y-g'(S)} \left(y-y\ws+\frac{\mu-r}{\gamma\sigma^2S}\right)S-(g(S)-g'(S)S)\right]\nonumber\\
&&=-\gamma\left[yS-g(S)- \sign{y-g'(S)}\frac{\mu-r}{\gamma\sigma^2}\varepsilon+O(\varepsilon^{\frac43})\right]\label{eq:Q-liab-final-cond}.
\end{eqnarray}

In this case, for $(S,y,T)\in \overline \NTw$ using the final time conditions $H_1\w(S,T)=\gamma g(S),~H_2\w(S,T)=0$ and $H_3{\wpm}(S,T) = \mp M_1\w$ and the boundedness of $H_4\w$ shown in Lemma \ref{lemma:bounds}, we also have that
\begin{equation}
Q^{(w)^{\pm}}(S,y,T)=\exp{\left\{-\gamma S y + \gamma g(S)  \mp\varepsilon M_1\w +O(\varepsilon^{\frac43})   \right\}}.
\label{eq:Q-liab-final-cond1}
\end{equation}
Comparing \eqref{eq:Q-liab-final-cond1} with exponentiated \eqref{eq:Q-liab-final-cond}, we conclude from the definition of $M^{\w}_1$ that for $\varepsilon>0$ small enough %
$\pm Q^{(w)^{\pm}}(S,y,T) \le \pm\exp{ \left\{ \mp\frac\epsilon2 M_1^{\w}\right\}  } Q^{\w}(S,y,T)\le \pm Q^{\w}(S,y,T) .$

For $j=1,w$ consider first the set $\left\{S\colon  y^{\jm}(S,T)\ge g'(S)\Ind\right\}$. On it we consider the following cases:
\newline
{\em Case $y \ge y^{\jp}$:} We have
\begin{eqnarray*}
&&\pm Q^{\jpm}(S,y,T)=\pm\exp{\{-\gamma(1-\varepsilon)S(y-y^{\jp}) \}}Q^{\jpm}(S,y^{\jp},T)\\
&& \le \pm\exp{\{-\gamma(1-\varepsilon)S(y-y^{\jp}) \}}Q^{\j}(S,y^{\jp},T)=\pm Q^{\j}(S,y,T),
\end{eqnarray*}
with the last equality readily follows from the final boundary condition \eqref{eq:Q-final-time}.\newline
{\em Case $g'(S)\Ind\le y \le y^{\jm}$:} %
We have  
\begin{eqnarray*}
&&\pm Q^{\jpm}(S,y,T)=\pm\exp{\{-\gamma(1+\varepsilon)S(y-y^{\jm}) \}}Q^{\jpm}(S,y^{\jm},T)\\
&&\le \pm\exp{\left\{-\gamma(1+\varepsilon)S(y-y^{\jm})\mp\frac12\varepsilon M_1^{\j}  \right\}}Q^{\j}(S,y^{\jm},T)\\
&&=\pm\exp{\left\{-\gamma(1+\varepsilon)S(y-y^{\jm})  \mp\frac12\varepsilon M_1^{\j}-\gamma S(1-\varepsilon) (y^{\jm}-y) \right\}} Q^{\j}(S,y,T)\\
&&=\pm\exp{\left\{ 2\gamma\varepsilon S (y^{\jm}-y)    \mp\frac12\varepsilon M_1^{\j}\right\}}Q^{\j}(S,y,T) \le  \pm Q^{\j}(S,y,T).
\end{eqnarray*}

Indeed, the last inequality follows from the fact that $S\abs{y^{\jm}-y}\le S(y^{\jstr}-g'(S)\Ind)=\frac{\mu-r}{\gamma\sigma^2}$, and the second equality follows from the fact that  $Q^{\j}(S,y,T)=\exp{\left\{\gamma(1-\varepsilon)S(y^{\jm}-y) \right\} } Q^{\j}(S,y^{\jm},T),$ for $y,S$ such that $g'(S)\Ind\le y \le y^{\jm}$, which in turn is a simple consequence of \eqref{eq:Q-final-time}.
\newline
{\em Case $y<g'(S)\Ind$:} We have
\begin{eqnarray*}
&&\pm Q^{\jpm}(S,y,T)=\pm\exp{\{-\gamma(1+\varepsilon)(y-g'(S)\Ind)S \}}Q^{\jpm}(S,g'(S)\Ind,T)\\
&&\le\pm\exp{\{-\gamma(1+\varepsilon)(y-g'(S)\Ind)S\}}Q^{\j}(S,g'(S)\Ind,T) =  \pm Q^{\j}(S,y,T).
\end{eqnarray*}
The sets $\left\{S\colon  y^{\jm}(S,T)<g'(S)\Ind\le y^{\jp}(S,T)\right\}$ and 
$\left\{S\colon  y^{\jp}(S,T)<g'(S)\Ind\ \right\}$ are treated similarly.

\subsection{ \emph{{Step 2: Verification that $\pm\D Q^{\jpm} \ge 0$ in $\overline\NT,~j=1,w$}}} \label{step2}
Using \eqref{eq:y2Y-func} inside $\NT$, we calculate that 
\begin{eqnarray}
Q_S^{\jpm}(S,y,t)&=& \left\{-\frac{\gamma  (\yjstr+ \epsilon^{\frac13} Y)}{\delta}+H_{0_S}^{\j}(S,t)+ \varepsilon^{\frac23} H_{2_S}^{\j}(S,t)\right.\label{eq:Q_S}\\
&+& \left.\varepsilon H_{3_S}^{\jpm}(S,t)+\varepsilon^{\frac43}H_{4_S}^{\j}(S,Y,t)-\epsilon \yjstr_SH_{4_Y}^{\j}(S,Y,t)\right\}Q^{\jpm}(S,y,t),\nonumber
\end{eqnarray}
\begin{eqnarray}
Q_{SS}^{\jpm}(S,y,t)&=& \left\{-\frac{\gamma  (\yjstr+ \epsilon^{\frac13} Y)}{\delta}+H_{0_S}^{\j}(S,t)+ \varepsilon^{\frac23} H_{2_S}^{\j}(S,t)\right.\label{eq:Q_SS}\\
&& \left.+\varepsilon H_{3_S}^{\jpm}(S,t)+\varepsilon^{\frac43}H_{4_S}^{\j}(S,Y,t)-\epsilon \yjstr_SH_{4_Y}^{\j}(S,Y,t)\right\}^2Q^{\jpm}(S,y,t)\nonumber\\
&+& \left\{H_{0_{SS}}(S,t) + \varepsilon^{\frac23} H_{2_{SS}}^{\j}(S,t) +\varepsilon H_{3_{SS}}^{\jpm}(S,t)-\epsilon \yjstr_{SS}H_{4_{Y}}^{\j}(S,Y,t) \   \right.\nonumber\\
&&\left.+\varepsilon^{\frac43}H_{4_{SS}}^{\j}(S,Y,t)-2\epsilon \yjstr_{S}H_{4_{YS}}^{\j}(S,Y,t) + \epsilon^{\frac23} \left(\yjstr_{S}\right)^2H_{4_{YY}}^{\j}(S,Y,t)\right\}%
Q^{\jpm}(S,y,t),\nonumber
\end{eqnarray}
\begin{eqnarray}
Q_t^{\jpm}(S,y,t)&=& \left\{\frac{r\gamma S (\yjstr+ \epsilon^{\frac13} Y)}{\delta}+H_{0_t}^{\j}(S,t)+ \varepsilon^{\frac23} H_{2_t}^{\j}(S,t)\right.\label{eq:Q_t}\\
&+& \left.\varepsilon H_{3_t}^{\jpm}(S,t)+\varepsilon^{\frac43}H_{4_t}^{\j}(S,Y,t)-\epsilon \yjstr_{t}H_{4_Y}^{\j}(S,Y,t)\right\}Q^{\jpm}(S,y,t).\nonumber
\end{eqnarray}
Using the boundedness of $S^2y^{\jstr}_S, SY^{\j},  H_4, SH_{2_S}^{(j)},SH_{4_S}^{(j)},Sy_S^{\jstr}H_{4_Y},$
$S^2y_S^{\jstr}H_{4_{YS}},$
$ S^2\left(y_S^{\jstr}\right)^2H_{4_{YY}},$  
$S^2H_{4_{SS}}^{(j)},$ 
$ S^2y_{SS}^{\jstr}H_{4_{Y}} $ that are shown in Remark \ref{remark:NT-width} and Lemma \ref{lemma:gamma-bounds}, using \eqref{eq:y2Y-func}, we calculate $\pm\D Q^{\jpm} $, on $\overline\NT,~j=1,w$  %
\begin{eqnarray}
&&\!\!\!\!\!\!\!\!\!\!\! \pm\D\left(Q^{\jpm}\right)= \pm Q^{\jpm}\label{eq:DQ-NT}\\
&&\!\!\!\!\!\!\!\!\!\!\!\quad\times\left[ \varepsilon^0 \left\{H_{0_t}^{\j}+\frac{r\gamma S\yjstr}{\delta}+\mu S\left(H_{0_S}^{\j}-\frac{\gamma\yjstr}{\delta}\right)+\frac{\sigma^2S^2}{2}\left(H_{0_S}^{\j}-\frac{\gamma\yjstr}{\delta}\right)^2\right.
+\frac{\sigma^2S^2}{2}H_{0_{SS}}^{\j}\right\}\nonumber\\
&&\!\!\!\!\!\!\!\!\!\!\!\qquad+\varepsilon^{\frac13}\left\{\frac{(r-\mu)\gamma SY}{\delta}-\frac{\sigma^2S^2\gamma Y}{\delta}\left(H_{0_S}^{\j}-\frac{\gamma\yjstr}{\delta}\right) \right\}\nonumber\\
&&\!\!\!\!\!\!\!\!\!\!\!\qquad+\varepsilon^{\frac23}\left\{H_{2_t}^{\j}+rSH_{2_S}^{\j}+\frac{\sigma^2S^2}{2}H_{2_{SS}}^{\j}+\frac{\sigma^2S^2}{2}\left(\yjstr_S\right)^2H_{4_{YY}}^{\j}+\frac{\gamma^2\sigma^2S^2Y^2}{2\delta^2}\right\}\nonumber\\
&&\!\!\!\!\!\!\!\!\!\!\!\qquad+\varepsilon\left\{H_{3_t}^{\jpm}-\yjstr_t H_{4_Y}^{\j}+\mu S\left(H_{3_S}^{\jpm}-\yjstr_S H_{4_Y}^{\j}\right)\right.\nonumber\\
&&\!\!\!\!\!\!\!\!\!\!\!\qquad\quad+\frac{\sigma^2S^2}{2}\left[H_{3_{SS}}^{\jpm}-y_{SS}^{*}H_{4_Y}^{\j}-2\yjstr_S H_{4_{YS}}^{\j} %
\left.\left.+2\left(H_{0_S}^{\j}-\frac{\gamma\yjstr}{\delta}\right)\left(H_{3_S}^{\jpm}-\yjstr_S H_{4_Y}^{\j}\right)-\frac{2\gamma Y}{\delta}H_{2_S}\right]\right\} + O(\varepsilon^{\frac43})\right]%
\ge 0\nonumber.
\end{eqnarray}
The $\varepsilon^{0}, \varepsilon^{\frac13}$ and $\varepsilon^{\frac23}$ terms are zero by construction of $H_0^{\j}, H_2^{\j}, H_4^{\j}$ and $\yjstr$. The coefficient of the $-\varepsilon H_{4_Y}^{\j}$ term is
\begin{eqnarray*}
&&\yjstr_t+\mu S \yjstr_S+\frac{\sigma^2S^2}{2}\yjstr_{SS}+\sigma^2S^2 \left(H_{0_S}^{\j}-\frac{\gamma \yjstr}{\delta}\right)\yjstr_S
=\yjstr_t+rS\yjstr_S+\frac{\sigma^2S^2}{2}\yjstr_{SS}\\
&&=\frac{\delta(\mu-r)}{\gamma S}+\left(V_{0_{St}}+rSV_{0_{SS}}+\frac{\sigma^2S^2}{2}V_{0_{SSS}}\right)\ind_{\{j=w\}}=\frac{\delta(\mu-r)}{\gamma S}-\sigma^2SV_{0_{SS}}\ind_{\{j=w\}}=-\sigma^2S\yjstr_S,
\end{eqnarray*}
where we have used \eqref{eq:y-star} and the fact that $V_{0_S}$ satisfies a Black-Scholes equation $V_{0_{St}}+(r+\sigma^2)SV_{0_{SS}}+\frac{\sigma^2S^2}{2}V_{0_{SSS}}=0$.
So it is enough to show that for $\varepsilon>0$ small enough, and for $j=1,w$ and $(S,y,t)\in \overline\NT$ 
$$
\pm\left(H_{3_t}^{\pm}+rSH_{3_S}^{\pm}+\frac{\sigma^2S}{2}H_{3_{SS}}^{\pm}\right)\ge \pm\sigma^2S\abs{\frac{S\gamma Y}{\delta}H_{2_S}-y_S^{\jstr}H_{4_Y}+Sy_S^{\jstr}H_{4_{YS}}}  \pm1,\nonumber
$$
which is shown in Lemma \ref{lemma:bounds}, where we have again used \eqref{eq:y2Y-func}.

$\hfill\Box$

\subsection{ \emph{{Step 3a: Verification that $\pm\D Q^{\jp} \ge 0$ inside $\B$ region, $j=1,w$}}} \label{step3a}
We have shown for $j=1,w$ that $Q^{\jp} \in C^{2,1,1}$, so from Step 2 it follows that $\D Q^{\jp}(S,y^{\jm},t)\ge 0,~(S,t)\in\Rpplus\times\zeroClosedTclosed$. Using \eqref{eq:D}, \eqref{eq:Q_S-in-B}, \eqref{eq:Q_t-in-B} and \eqref{eq:Q_SS-in-B} the reader can verify that to conclude that $\D Q^{\jp} \ge 0$ inside $\B$ region it is enough to show for $j=1,w$ that 
\begin{eqnarray*}
&& (r-\mu)S\frac{\gamma(1+\varepsilon)}{\delta}(y-y^{\jm})Q^{\jp}-2\frac{\sigma^2S^2}{2}\left(Q_S^{\jp}+
Q_y^{\jp}y_S^{\jm}\right)\frac{\gamma(1+\varepsilon)}{\delta}(y-y^{\jm})\\
&&+\frac{\sigma^2S^2\gamma^2(1+\varepsilon)^2}{2\delta^2}Q^{\jp}\left(y-y^{\jm}\right)^2-2\frac{\sigma^2S^2}{2}Q^{\jp}\frac{\gamma^2(1+\varepsilon)^2S}{\delta^2}y_S^{\jm}(y-y^{\jm})\ge 0.
\end{eqnarray*}
Here and in the rest of Step 3a, $Q^{\jp}$ and its derivatives are evaluated at $(S,y^{\jm},t)$. We rewrite the inequality above as
\begin{eqnarray*}
&&-\frac{\gamma(1+\varepsilon)S}{\delta}(y-y^{\jm})\left[(\mu-r)Q^{\jp}+\sigma^2SQ_S^{\jp}\right]+\frac{\gamma^2(1+\varepsilon)^2\sigma^2S^2}{2\delta^2}\left(y-y^{\jm}\right)^2Q^{\jp}\\
&&-\frac{\sigma^2S^2\gamma(1+\varepsilon)}{\delta}(y-y^{\jm})y_S^{\jm}\left[\frac{\gamma(1+\varepsilon)S}{\delta}Q^{\jp}+Q_y^{\jp}\right]\ge0.
\end{eqnarray*}
The second term is non-negative and third term is zero by \eqref{eq:equality-Q_y}, so it is enough to show that
\begin{equation}
-\frac{\gamma(1+\varepsilon)\sigma^2S^2}{\delta}(y-y^{\jm})\left[\varepsilon^{\frac13}\frac{\gamma Y^{\j}}{\delta}+\varepsilon^{\frac23}H_{2_S}^{\j}+\varepsilon H_{3_S}^{\jp}+\varepsilon^{\frac43}H_{4_S}^{\j}-\varepsilon \yjstr_SH_{4_Y}^{\j} \right]Q^{\jp}
\label{eq:step3-inequality}
\end{equation}
is non-negative. Here, we have used \eqref{eq:Q_S} and \eqref{eq:y-star}.
By Remark \ref{remark:NT-width} and Lemma \ref{lemma:bounds} the terms in the square brackets scaled by $S$ are bounded. Since $SY^{\j}>0$ and is bounded away from zero, we conclude that \eqref{eq:step3-inequality} is non-negative for $\varepsilon$ small enough for both $j=1,w$.

$\hfill\Box$

\subsection{ \emph{{Step 3b: Verification that $\D Q^{\jp} \ge 0$ inside $\S$ region, $j=1,w$}}} \label{step3b}
The proof is similar to proof of Step 3a, and is omitted for brevity.
\subsection{ \emph{{Step 4a: Verification that $\pm Q_y^{\jp}\pm\frac{\gamma(1\pm\varepsilon)S}{\delta} Q^{\jp} \ge 0$ in $\overline\NT$, $j=1,w$}}} \label{step4a}
From equation \eqref{eq:Q_y-in-NT} 
$\pm Q_y^{\jp}\pm\frac{\gamma(1\pm\varepsilon)S}{\delta} Q^{\jp} = \varepsilon\left[ \frac{\gamma S} {\delta} \pm H_{4_Y}^{\j} \right]Q^{\jp},$ which is non-negative in $\overline\NT$ by Remark \ref{remark:H_4_Y}.
\subsection{ \emph{{Step 4b: Verification that $Q_y^{\jp} +\frac{\gamma(1+\varepsilon)S}{\delta} Q^{\jp} \ge 0$ and $-Q_y^{\jp} -\frac{\gamma(1-\varepsilon)S}{\delta} Q^{\jp} \ge 0$ inside $\S$ and $\B$ regions respectively, $j=1,w$}}} \label{step4b}
Inside $\S$ region $Q_y^{\jp} +\frac{\gamma(1+\varepsilon)S}{\delta} Q^{\jp} \ge Q_y^{\jp} +\frac{\gamma(1-\varepsilon)S}{\delta} Q^{\jp}=0 $, because $Q^{\jp}\ge0$. The proof inside $\B$ is similar.

\subsection{ \emph{{Step 5: Conclusion}}} This concludes the proof of Theorem \ref{Thm1}, as we have shown the smoothness of $Q^{\jpm}$ in Lemmas \ref{lemma:Q_C1} and \ref{lemma:Q_C2}. The final time condition was proved in Step 1. Steps 2-4 show that $\mathcal H_3 Q^{\jp} \ge 0$ in $\Rpplus\times\R\times\zeroClosedTclosed$. 

In Step 2 we showed that $\mathcal H_3 Q^{\jm} \le 0$ in $\NT$ . Since inside $\S$ we have $\mathcal H_3 Q^{\jm} \le Q_y^{\jm} +\frac{\gamma(1-\varepsilon)S}{\delta} Q^{\jm}=0$ and similarly inside $\B$ we have that $\mathcal H_3 Q^{\jm} \le0$. We conclude from smoothness of $Q^{\jm}$ that $\mathcal H_3 Q^{\jm} \le 0$ in $\Rpplus\times\R\times\zeroClosedTclosed$. 

$\hfill\Box$

\section{``Nearly-Optimal" Strategy}\label{sec:nearly-optimal_strat}

In this section we will prove existence of the strategy $(\tilde L^{\j}, \tilde M^{\j}),~j=1,w$, associated with the {\em``no-trade"}, buy and sell regions $\NT, \B$ and $\S$ from \eqref{eq:NT}, \eqref{eq:buy} and \eqref{eq:sell}. Moreover, in Section \ref{sec:comp} we will prove Theorem \ref{thm:comp} -- the verification argument, from which the result of Theorem \ref{thm:Thm2} that $(\tilde L^{\j}, \tilde M^{\j})$ is a ``nearly optimal" strategy would follow. The strategy $(\tilde L^{\j}, \tilde M^{\j})$ requires trading anytime the position is inside the buy or sell regions until the position reaches the boundary of the $\overline\NT$ region. Then the strategy calls for buying (respectively selling) stock whenever the position is on the boundary of $\B$ (respectively $\S$), so that agent's position does not leave $\overline\NT$. On the boundaries of the {\em``no-trade"} region these trades increase $\tilde L^{\j}$ or $\tilde M^{\j}$ and push the associated process $(s, \tilde B^{\j}_s, \y^{\j}_s, S_s)$ in direction pointing to the inside of $\NT$. We refer to the process $(s, \tilde B^{\j}_s, \y^{\j}_s, S_s)$ as reflected process and to these directions as directions of reflection.

For $j=1,w$ define the {\em``no-trade"} region in four variables as 
\begin{equation}
\NT_4 \define \left\{(t,B,yS,S) | (S,y,t)\in \NT, B\in \R\right\}. 
\label{eq:NT3}
\end{equation}
Similarly we define the buy and sell regions  $\B_4, \S_4$.

{\sc Proof of Lemma \ref{lemma:existence}:}
For $j=1,w$, we define the strategy $(\tilde L^{\j},\tilde M^{\j})$ to be the trading strategy associated with with the {\em``no-trade"}, buy and sell regions $\NT_4, \B_4$ and $\S_4$. That is let the process $(s, \tilde B^{\j}_s, \y^{\j}_s, S_s)\big|_{s\in\tClosedTclosed}$ be given by \eqref{eq:dB} - \eqref{eq:dS}, such that $B_{t-}=B, y_{t-}=y,S_{t}=S$ and $(s, B_s, y_sS_s,S_s)\in\overline\NT_4,~t\le s\le T,$ so that the directions of reflections are $(0,-(1+\varepsilon),1,0)$ and $(0,1-\varepsilon,-1,0)$ on the buy and sell boundaries respectively. The existence of the trading strategy and the reflected process, by itself a non-trivial problem as the boundaries of the $\overline\NT_4$ region are not constant, is proved in Burdzy, Kang \& Ramanan \cite{BurdzyKangRamanan}. To apply their result, we need to use the fact that by Remark \ref{remark:NT-width} the width of the $\overline\NT_4$ region in $yS$ is bounded and is bounded away from zero. Moreover, on the set $\inf\left\{s\ge t \colon S_s=0\right\}=\infty$ of probability $1$, the process $(S_s, \y^{\j}_s, s)\big|_{s\in\tClosedTclosed}$ is a reflected process in $\NT$ region, $j=1,w$.

Our goal is now to show that 
\begin{equation}
\E_t^{B,y,S}\left[   \exp{ \left\{  -\gamma \left( \tilde B^{\j}_T + \y^{\j}_TS_T - V_0(S,T)\ind_{\{j=w\}}  \right)\right\}  } \right] <\infty
\label{eq:integrability}
\end{equation}
for $\varepsilon$ small enough. 
Solving \eqref{eq:dB} -- \eqref{eq:dS}, we find for $j=1,w$ that
\begin{eqnarray}
&&\!\!\!\!\!\!\!\tilde B^{\j}_T= \e{ r(T-t)}B-(1+\varepsilon)\int_t^T \e{r(T-s)}S_sd\tilde L^{\j}_s+(1-\varepsilon)\int_t^T \e{r(T-s)}S_sd\tilde M^{\j}_s,\label{eq:B_T}\\ %
&&\!\!\!\!\!\!\!\y^{\j}_TS_T-V_0(S_T,T)\ind_{\{j=w\}}=   \e{r(T-t)}yS-\e{r(T-t)}V_0(S,t)\ind_{\{j=w\}}\label{eq:yS_T}\\
&&%
-\int_t^T\e{r(T-s)}\left(-rV_{0}(S_s,s)+V_{0_t}(S_s,s)+\frac12\sigma^2S_s^2V_{0_{SS}}(S_s,s)\right)ds\ind_{\{j=w\}} \nonumber\\
&&+\int_t^T \e{r(T-s)}S_s\left(d\tilde L^{\j}_s-d\tilde M^{\j}_s\right)+\int_t^T \e{r(T-s)}\left(\y^{\j}_s-V_{0_S}(S_s,s)\ind_{\{j=w\}}\right)dS_s \nonumber\\
&&-r\int_t^T \e{r(T-s)}\y^{\j}_sS_sds= \e{r(T-t)}yS-\e{r(T-t)}V_0(S,t)\ind_{\{j=w\}}\nonumber\\
&&-r\int_t^T \e{r(T-s)}\left(\y^{\j}_sS_s -S_sV_{0_S}(S_s,s)\ind_{\{j=w\}}\right)ds+\int_t^T \e{r(T-s)}S_s\left(d\tilde L^{\j}_s-d\tilde M^{\j}_s\right)\nonumber\\
&&+\int_t^T \e{r(T-s)}\left(\y^{\j}_s-V_{0_S}(S_s,s)\ind_{\{j=w\}}\right)dS_s,\nonumber
\end{eqnarray} 
where we have used the Black -Scholes equation \eqref{eq:BS}. Using the fact that $\y^{\j}_sS_s-S_sV_{0_S}(S_s,s)\ind_{\{j=w\}}$ is bounded and that 
\begin{eqnarray*}
&&\E_t^{B,y,S}\left[ \exp{ \left\{ \sigma\int_t^T \e{r(T-s)}\left(\y^{\j}_s-V_{0_S}(S_s,s)\ind_{\{j=w\}}\right)S_s dZ_s  \right\} } \right]\\
&&= \E_t^{B,y,S}\left[ \exp{ \left\{\frac{ \sigma^2}{2}\int_t^T \e{2r(T-s)}\left(\y^{\j}_s-V_{0_S}(S_s,s)\ind_{\{j=w\}}\right)^2S_s^2 ds  \right\} } \right]
\end{eqnarray*}
is finite, 
We conclude that to prove \eqref{eq:integrability}, it is enough to show that $\E_t^{B,y,S}\left[   \exp{ \left\{  \gamma\varepsilon  \int_t^T S_s\left(d\tilde L^{\j}_s+d\tilde M^{\j}_s\right)\right\}  } \right] <\infty$ for $\varepsilon$ small enough. 

Define $\hat X_s\define \left(\y^{\j}_s-\yjstr\right)S_s$. Using \eqref{eq:y-star} we compute 
\begin{eqnarray*}
d\hat X_s&=&d\left(\y^{\j}_sS_s  -\frac{\delta(\mu-r)}{\gamma \sigma^2}  -S_sV_{0_S}\ind_{\{j=w\}}\right) \\
 &=&S_sd\left(\tilde L^{\j}_s-\tilde M^{\j}_s\right)+ \y^{\j}_sdS_s - \frac{r\delta(\mu-r)}{\gamma \sigma^2}ds\\
&&\quad-\left( S_sV_{0_{St}}ds + \left[V_{0_S} +S_sV_{0_{SS}}\right]dS_s+ \sigma^2S_s^2\left[\frac12S_sV_{0_{SSS}}+V_{0_{SS}}  \right]ds   \right)\ind_{\{j=w\}}.
\end{eqnarray*}
Due to reflections $\left(\hat L^{\j}_\tau,\hat M^{\j}_\tau\right)\define\left(\int_t^\tau S_sd\tilde L^{\j}_s,\int_t^\tau S_sd\tilde %
M^{\j}_s\right)$, the process $\hat X_s$ remains inside $[-\varepsilon^{\frac13} S_sY^{\j}, \varepsilon^{\frac13} S_sY^{\j}]$, $s\in\tClosedTclosed$. 
Its drift and variance are
\begin{eqnarray}
\hat\alpha_s&=&- \frac{r\delta(\mu-r)}{\gamma \sigma^2} + \mu S_s\left(\y^{\j}_s -V_{0_S} \ind_{\{j=w\}}\right) %
-  \left( S_sV_{0_{St}} +  \mu S_s^2V_{0_{SS}}+ \sigma^2S_s^2\left[\frac12S_sV_{0_{SSS}}+V_{0_{SS}}  \right]  \right)\ind_{\{j=w\}},\label{eq:drift}\\
\hat\sigma_s&=&\sigma \left(\y^{\j}_s -   \left[ V_{0_S} +S_sV_{0_{SS}} \right] \ind_{\{j=w\}} \right)S_s\label{eq:vol}
\end{eqnarray}
are bounded because of Lemma \ref{lemma:gamma-bounds} and Remark \ref{remark:NT-width}. Moreover, for $\varepsilon$ small engouh the volatility $\hat\sigma_s\ge\frac{\sigma}2\varepsilon_1$ is bounded away from zero by Assumption \ref{A2}.
From Remark \ref{remark:NT-width} it follows that there exists a constant $c>0$ such that $[-c,c]\subset \cap_{j=1,w}[- S_sY^{\j},  S_sY^{\j}]$, $\forall s \in \tClosedTclosed$. Using the comparison result of Burdzy, Kang \& Ramanan \cite{BurdzyKangRamanan} it is sufficient to show the finiteness of $\E_t^{B,y,S}\left[   \exp{ \left\{ \gamma\varepsilon  \int_t^T S_s\left(d\overline L_s+d\overline M_s\right)\right\}  } \right] $, where $(\overline L,\overline M)$ are the reflecting processes of the reflected process $d\overline X_s\define\overline X_s(\hat \alpha_s ds+\hat \sigma_s dZ_s)+\overline L_s-\overline M_s$, starting from $\overline X_{t-}\define (y-\yjstr) S$ that remains inside $[-c\varepsilon^{\frac13},c\varepsilon^{\frac13}]$. This is true for $\varepsilon$ small enough by Lemma 5.8 of Jane\v{c}ek \& Shreve, \cite{JanecekShreve2}. Specifically, let $\gamma_3,\gamma_4$ be the constants appearing in the Lemma 5.8, then the assertion is true for any $\varepsilon>0$ satisfying $\varepsilon\le \gamma_3\wedge2\gamma_4c\varepsilon^{\frac13}$.

To finish this proof, we need to show the admissibility of the strategy $(\tilde L^{\j},\tilde M^{\j})$ for $j=1,w$. We already know that $\left(\y^{\j}_s-\yjstr\right)S_s,~s\in\tClosedTclosed$ is bounded, so we are left to prove that 
\begin{equation}
\left\{ \exp{ \left\{  -\gamma \left( \tilde B^{\j}_\tau + \y^{\j}_\tau S_\tau - V_0(S,\tau)\ind_{\{j=w\}}  \right)\right\}  } \right\}\Big|_{\tau\in\mathcal T_T}
\label{eq:UI}
\end{equation}
is uniformly integrable for $\varepsilon$ small enough. By the argument above $\forall \tau\in\mathcal T_T$
\begin{equation}
\E_t^{B,y,S}\left[ \exp{ \left\{  -\gamma \left( \tilde B^{\j}_\tau + \y^{\j}_\tau S_\tau - V_0(S,\tau)\ind_{\{j=w\}}  \right)\right\} }  \right] \le C \E_t^{B,y,S}\left[ \exp{ \left\{  \gamma\varepsilon  \int_t^T S_s\left(d\tilde L^{\j}_s+d\tilde M^{\j}_s\right)\right\}  } \right],
\label{eq:UI1}
\end{equation}
where $C$ is a constant. The uniform integrability of \eqref{eq:UI} follows from the integrability of the right hand side of \eqref{eq:UI1}.

$\hfill\Box$

\section{Proof of Theorem \ref{thm:comp}:}\label{sec:comp}

We are now ready to proof the Theorem \ref{thm:comp}. As a reminder, we have defined $\psi^{\jpm}$ in \eqref{eq:psi} as
\begin{equation}
\psi^{\jpm}(t,B,y,S) \define\Util-\exp{\left\{-\gamma\frac{B}{\delta}\right\}}Q^{\jpm}(S,y,t),~j=1,w.
\label{eq:psi1}
\end{equation}
Fix $j\in\left\{1,w\right\}$, and let $0\le t< t_1\le T$, and fix a starting point at time $t$ to be $(B, y, S)\in\R\times\R\times\Rpplus$.  Let $(L_s,M_s)\big|_{s\in\tClosedtOneClosed}$ be an admissible trading strategy, defining the process $(B_s, y_s, S_s),~s\in[t,t_1]$ be the process defined by \eqref{eq:dB}--\eqref{eq:dS}.
Define $\tau_n=t_1\wedge \inf\{t \le s \le t_1; B_s+c(y_s,S_s)- V_0(S_s,s)\ind_{\{j=w\}} \le -n\}$ then $\lim\limits_{n\rightarrow\infty}\tau_n= t_1$ a.s..

Since $\psi^{\pm}\in C^{1,1,1,2}\left(\zeroClosedTclosed\times\R\times\R\times\Rpplus\right) $, we can apply It\^o's rule to get 
\begin{eqnarray}
&&\psi^{\jpm}(\tau_n, B_{\tau_n}, y_{\tau_n},S_{\tau_n})-\psi^{\jpm}(t, B, y,S)  = -\int_t^{\tau_n} 
\Bigl[\mathcal L \psi^{\jpm}(s,B_s,y_s,S_s)\,\du s\nonumber\\
&& \qquad+ \big((1+ \varepsilon)S_s\psi^{\jpm}_B(s, B_s,y_s,S_s) -\psi^{\jpm}_y(s, B_s,y_s,S_s)
\big)\,\du L_s  \nonumber\\
&& \qquad +\big(-(1-\varepsilon)S_s\psi^{\jpm}_B(s, B_s,y_s,S_s) +\psi^{\jpm}_y(s, B_s,y_s,S_s)\big)\,\du M_s\Bigr] %
+ \int_t^{\tau_n}\psi^{\jpm}_S(s, B_s,y_s,S_s)S_s \, \sigma\,\du W_s, %
\label{4.5}
\end{eqnarray}
where $\mathcal L$ is the diffusion generator from \eqref{eq:L}.
From Theorem \ref{Thm1} using Remark \ref{remark:viscsol-WW} it follows that
the $ds, dL_s$ and $dM_s$ integrands in \eqref{4.5} are  
are all non-negative. Thus $\psi^{\jp}$ is a (local) supermartingale.

We next show that $\psi^{\jp}(s, B_s,y_s,S_s)$ is a (true) supermartingale. %
From Lemma \ref{lemma:bounds} both $H_2^{\j}$ and $H_4^{\j}$ are bounded. In case $(s,B_s,y_s,S_s)\in\overline\NT_4$ we have
\begin{eqnarray*}
&&\!\!\!\!\!\!\!\!\!\!\!\!-\frac{\gamma}{\delta}(B_s+y_sS_s)+H_0^{\j}(S_s,s)+\varepsilon H_3^{\jp}(S_s,s)=-\frac{\gamma}{\delta}(B_s+y_sS_s)-\frac{(\mu-r)^2(T-s)}{2\sigma^2}
\label{eq:uniform-intgrl1}\\
&&\!\!\!\!\!\!\!\!\!\!\!\!+\frac{\gamma}{\delta}V_0(S_s,s)\ind_{\{j=w\}}%
- \varepsilon M^{\j}(T-s)- \varepsilon M^{\j}_1\le-\frac{\gamma}{\delta}\left(B_s+y_sS_s-V_0(S_s,s)\ind_{\{j=w\}}\right).\nonumber
\end{eqnarray*}
It follows that 
\begin{eqnarray*}\psi^{\jp}(s,B_s,y_s,S_s)&=&-\exp{ \left\{-\frac{\gamma}{\delta}(B_s+y_sS_s) + H_0^{\j}(S_s,s) +\varepsilon H_3^{\jp}(S_s,s) + O(\varepsilon^{\frac23}) \right\} }\\
&\ge& - \exp { \left\{-\frac{\gamma}{\delta}\left(B_s+y_sS_s-V_0(S_s,s)\ind_{\{j=w\}} + O(\varepsilon^{\frac23})\right)\right\} }.
\end{eqnarray*}
In case $(s,B_s,y_s,S_s)\in\overline\B_4$ from \eqref{eq:Q_buy} we have
\begin{eqnarray} 
&&\psi^{\jp}(s,B_s,y_s,S_s)=\Util-\exp{\left\{-\frac{\gamma}{\delta}B_s \right\}}Q^{\jp}(S_s,y_s,s)\label{eq:uniform-intgrl2}\\
&&= \Util-\exp{\left\{-\frac{\gamma}{\delta}\left(B_s +(1+\varepsilon)\left(y_s-y^{\jm}_s \right)S_s\right) \right\}}Q^{\jp}(S_s,y^{\jm}_s,s)   \nonumber\\
&&=\Util-\exp{\left\{-\frac{\gamma}{\delta}\left(B_s +(1+\varepsilon)y_sS_s -\varepsilon y^{\jm}_sS_s  \right) \right\}}%
\exp {\left\{ H_0^{\j}(S_s,s) +\varepsilon H_3^{\jp}(S_s,s) + O(\varepsilon^{\frac23}) \right\}}\nonumber\\
&&=\Util-\exp{\left\{-\frac{\gamma}{\delta}\left(B_s +y_sS_s +\varepsilon (y_s-y^{\jm}_s)S_s  -V_0(S_s,s))\ind_{\{j=w\}}  + O(1)\right)  \right\}}\nonumber\\
&&=\Util-\exp{\left\{-\frac{\gamma}{\delta}\left(B_s +y_sS_s -V_0(S_s,s))\ind_{\{j=w\}}  + O(1)\right)  \right\}},\nonumber
\end{eqnarray} 
where in the last equality we have used the facts that $\abs{y_s-y^{\jm}_s}S_s \le\abs{y_s-y^{\jstr}_s} S_s + \abs{y^{\jstr}_s-y^{\jm}_s}S_s$ is bounded by admissibility of the strategy and Remark \ref{remark:NT-width}.
We get a similar inequality when $(s,B_s,y_s,S_s)\in \overline\S_4$.

Since $\left\{ \e{-\frac{\gamma}{\delta}\left(B_\tau+y_\tau S_\tau-V_0(S_\tau,\tau)\ind_{\{j=w\}}\right)}\right\}\Big|_{\tau\in\mathcal T_{t_1}}$ is uniformly integrable, by Definition \ref{def:admis} of an admissible strategy, it follows that so is 
$\left\{ \psi^{\jp}(\tau, B_\tau, y_\tau,S_\tau)\right\}\Big|_{\tau\in\mathcal T_{t_1}}$. We conclude that $\psi^{\jp}$ is a supermartingale.

Now consider the ``nearly-optimal" strategy $(\tilde L^{\j}, \tilde M^{\j})$ and the reflected process that it generates $(s, \tilde B_s,\y_s,S_s),$ $s\in[t,t_1]$. We assert that $\psi^{\jm}(s,\tilde B_s,\y_s,S_s)$ is a submartingale. We apply It\^o's rule $\psi^{\jm}(s,\tilde B_s,\y_s,S_s)$ to get an equation similar to \eqref{4.5}. In this equation, by construction of the ``nearly optimal" policy the terms containing the integrals with respect to $d\tilde L^{\j}$ and $d\tilde M^{\j}$ are zero, and $\mathcal L \psi^{\jm}\le0$. We conclude that $\psi^{\jm}(s,\tilde B_s,\y_s,S_s)$ is a (local) submartingale, but since it is also non-positive, it is (true) submartingale.

$\hfill\Box$

\section{Conclusion}\label{sec:conclusion}

In this paper we have found an asymptotic power expansion of the price the contingent claim liability in the presence of proportional transaction costs $\varepsilon>0$ using utility indifference pricing. The expansion in powers of $\varepsilon^\frac13$ was done around the Black-Scholes price of the contingent claim in a model without transaction costs. To achieve this goal two problems of utility optimization have been solved. The first one -- utility optimization of the final wealth with the contingent claim liability, and the second one -- without the contingent claim liability. A lower and an upper bounds on the value function were constructed in each case. These bounds matched at the order $\varepsilon^\frac23$, hence the value functions were determined at order $\varepsilon^\frac23$. Finally, in both cases, a simple ``nearly optimal" strategy was constructed, whose utility asymptotically matches the leading terms of the appropriate value function.

\appendix
\numberwithin{equation}{section}

\section{Appendix}\label{sec:appendix}
\setcounter{theorem}{0}
\setcounter{equation}{0}

{\sc Proof of Lemma \ref{lemma:gamma-bounds}:}
Another way to express $V_0(S,t)$ is to use the risk neutral probability $\tilde \P$ in the complete market, with zero transaction costs. Then the Black-Scholes price $V_0(S,t)$ of the contingent claim is
\begin{equation}
V_0(S,t)=\e{-r(T-t)}\tilde \E_t^S\left[g(S_T)\right]=\e{-r(T-t)}\tilde \E_t^S\left[g(S\e{\sigma(Z_T-Z_t)+(r-\frac{\sigma^2}{2})(T-t) })\right].
\label{eq:V-rsik-neutral}
\end{equation}
From Assumption \ref{A1} $g$ is non-negative, it follows that $V_0\ge0$. Moreover, for $i\ge2$ we calculate 

$$ \frac{\partial^i V_0(S,t)}{\partial S^i}=\e{-r(T-t)} \tilde \E_t^S\left[\left(\e{\sigma(Z_T-Z_t)+(r-\frac{\sigma^2}{2})(T-t)}\right)^ig^{(i)}(S\e{\sigma(Z_T-Z_t)+(r-\frac{\sigma^2}{2})(T-t)} )\right],
$$
where by $g^{(i)}$ we mean the $i^{\mbox{th}}$ derivative.
We conclude that
\begin{equation}
\abs{S^i \frac{\partial^i V_0(S,t)}{\partial S^i}}=\e{-r(T-t)}\abs{\tilde \E_t^S\left[S_T^ig^{(i)}(S_T)\right]} \le \sup_{x\in\Rpplus} \abs{x^ig^{(i)}(x)},\label{eq:g-deriv-bound}
\end{equation}
By Assumption \ref{A1} the bound in \eqref{eq:g-deriv-bound} is finite and in a similar manner we conclude that $\abs{V_0-SV_{0_S}},~S^2\abs{V_{0_{SS}}},$
$ \abs{S^3V_{0_{SSS}}}$ and $S^4\abs{V_{0_{SSSS}}}$ are bounded on $\Rpplus\times\zeroClosedTclosed$ by $\sup_{ x>0} \abs{g(x)-g'(x)x},$ $\sup_{x>0}x^2\abs{g''(x)}$,  $\sup_{x>0}\abs{x^3g^{(3)}(x)}$ and $\sup_{x>0}x^4\abs{g^{(4)}(x)}$ respectively. 

We also calculate 
\begin{equation}
y_S^{\jstr}=\frac\delta\gamma H^{(j)}_{0_{SS}}-\frac{\delta(\mu-r)}{\gamma S^2\sigma^2}, ~j=1,w,
\label{eq:y_S-star}
\end{equation}
and from the definition of $H_0^{\j}$ in \eqref{eq:H_0} conclude that $S^2y_S^{\jstr}$ is bounded in $(S,t)\in\Rpplus\times\zeroClosedTclosed$ in both cases with and without contingent claim liability. Similar calculation shows that $S^3y_{SS}^{\jstr},S^4y_{SSS}^{\jstr}$ are also bounded in $(S,t)\in\Rpplus\times\zeroClosedTclosed$ for $j=1,w$. 

Moreover, from \eqref{eq:BS} we have that
\begin{equation}
V_{0_{SSt}} +\frac{\sigma^2S^2}{2}V_{0_{SSSS}} +(r+2\sigma^2)SV_{0_{SSS}}+(r+\sigma^2)V_{0_{SS}}=0.
\label{eq:BS_t}
\end{equation}
From the boundedness of $S^2V_{0_{SS}}, ~S^3V_{0_{SSS}}$ and $S^4V_{0_{SSSS}}$ on $\Rpplus\times\zeroClosedTclosed$ the boundedness of $S^2V_{0_{SSt}}$ also follows, and we conclude that $S^2y_{St}^{\jstr}=S^2V_{0_{SSt}}-\frac{r\delta(\mu-r)}{\gamma \sigma^2}$ is bounded in $(S,t)\in\Rpplus\times\zeroClosedTclosed$ in both cases with and without liability.

$\hfill\Box$

{\sc Proof of Lemma \ref{lemma:bounds}:}

From \eqref{eq:H_4} we see that $\abs{H\ju_4(S,Y,t)} \le \frac12\left(Y^{\j}\right)^2\left(\frac{3\gamma^2S}{2\delta^2y_S^{\jstr}} \right)^{\frac23} +\frac{\gamma^2\left(Y^{\j}\right)^4}{12\delta^2{y_S^{\jstr}}^{2}},$ for $(S,Y,t)\in\overline\NTY,~j=1,w.$ We use the estimates in Lemma \ref{lemma:gamma-bounds} and Remark \ref{remark:NT-width} to conclude that that $H_4\ju$ is bounded in $\overline\NTY$, $j=1,w$.

Similarly, using \eqref{eq:H_4_Y} -- \eqref{eq:H_4_t}, Lemma \ref{lemma:gamma-bounds} and Remark \ref{remark:NT-width} we can show that the terms $\abs{H\ju_{4_{YS}}}$, $\abs{Sy_S^{\jstr}H\ju_{4_{Y}}}$, $\abs{S^2y_S^{\jstr}H\ju_{4_{YS}}}$, $\abs{H\ju_{4_t}}$, $\abs{SH\ju_{4_S}}$, $\abs{S^2H\ju_{4_{SS}}}$,   $\abs{S^2y_{SS}^{\jstr}H\ju_{4_{Y}}}$ and  $\abs{S^2\left(y_S^{\jstr}\right)^2H\ju_{4_{YY}}}$are all bounded in their respective domains in both cases $j=1,w$.

In case of no contingent claim liability, $j=1$ then from Remark \ref{remark:y-star} it follows that $-\frac12\left(\frac{3\gamma^2S^4\sigma^3 \left(y_S^{\ones}\right) ^2}  {2\delta^2} \right)^{\frac23} = -\frac12 \left(\frac{3(\mu-r)^2}{2\sigma}\right)^\frac23.$ It follows that the solution to \eqref{eq:H_2} in case $j=1$ is given by
\eqref{eq:H_2-sol} as claimed.
Thus $H_2^{\one}$ is bounded and $H_{2_S}^{\one}=0$. We conclude that $ H_3^{\onepm}(S,t)=\mp M^{\one}(T-t)\mp M_1^{\one}$, as defined in \eqref{eq:H_3-def}, will satisfy \eqref{eq:H_3} with $M^{\one}$ big enough and  $M_1^{\one}$ given by \eqref{eq:M_1}.

We now concentrate on the case of holding a contingent claim liability $j=w$. 
We rewrite equation \eqref{eq:H_2} as $H_{2_t}^{(j)}+rSH_{2_S}^{(j)}+\frac{\sigma^2S^2}{2}H_{2_{SS}}^{(j)}=-f(S,t)$, where $f(S,t)=\frac12\left(\frac{3\gamma^2S^4\sigma^3\left({y^{\ws}_S}\right)^{2}}{2\delta^2} \right)^{\frac23}$, with the final boundary condition $H_2^{\w}(S,T)=0$. From Lemma \ref{lemma:gamma-bounds} we have that $f(S,t)$ and $Sf_S(S,t)$ are both bounded in $\Rpplus\times\zeroClosedTclosed$. Using the standard change of variables $S=\e{x},~\tau=\frac{\sigma^2}{2}(T-t)$ and $k=\frac{2r}{\sigma^2}$ let $\H_2(\tau,x) \define\e{\frac12(k-1)x+\frac14(k+1)^2\tau-k\tau}H_2^{\w}(S,t)$.  We see that $\H_2$ satisfies a non- homogeneous heat equation:
\begin{equation}
\H_{2_\tau}(\tau,x)=\H_{2_{xx}}(\tau,x)+\frac2{\sigma^2}\e{\frac12(k-1)x+\frac14(k+1)^2\tau-k\tau}f(S,t),
\label{eq:tilde-H_2_t-tmp}
\end{equation}
with zero initial time condition. The solution to this heat equation for $(\tau,x)\in[0,\frac{\sigma^2}{2}T]\times\R$ is
\begin{equation}
\H_2(\tau,x)=\frac2{\sigma^2}\int_0^\tau\int_{-\infty}^{\infty}\Psi(x-y,\tau-s)\e{\frac12(k-1)y+\frac14(k+1)^2s-ks}f(\e{y},T-\frac2{\sigma^2}s)dyds,
\label{eq:heat-eq-sol1-tmp}
\end{equation}
where $\Psi(x,t)=\frac1{\sqrt{4\pi t}}\exp{\left\{-\frac{x^2}{4t}\right\}}$ is the heat kernel. From the boundedness of $f$ it follows that %
$\H_2(\tau,x)=O\left(\e{\frac12(k-1)x}\right)$, and we conclude that $H_2^{\w}$ is bounded on $\Rpplus\times\zeroClosedTclosed$. 

Moreover, $\H_{2_x}$ also satisfies a non-homogeneous heat equation:
\begin{equation}
\H_{2_{x\tau}}(\tau,x)=\H_{2_{xxx}}(\tau,x)+\frac2{\sigma^2}\e{\frac12(k-1)x+\frac14(k+1)^2\tau-k\tau}\left(\frac12(k-1)f+\e{x}f_S\right)(S,t),
\label{eq:tilde-H_2_x-tmp}
\end{equation}
with zero initial condition. Using the fact that $\frac{\partial x}{\partial S} = \e{ -x}$ we calculate that
$$H_{2_S}^{\w}(S,t)=\e{-\frac12(k-1)x-\frac14(k+1)^2\tau+k\tau } \e{ -x} \left(-\frac12(k-1)\tilde H_2+\H_{2_x}\right)(\tau,x).$$
Let $\tilde\H_2=-\frac12(k-1)\H_2+\H_{2_x}$. Then $H_{2_S}^{\w}(S,t)=\e{-\frac12(k+1)x-\frac14(k+1)^2\tau+k\tau}\tilde\H(\tau,x)$, and $\tilde\H_2$ satisfies a non-homogeneous heat equation:
\begin{equation}
\tilde\H_{2_\tau}(\tau,x)=\tilde\H_{2_{xx}}(\tau,x)+\frac2{\sigma^2}\e{\frac12(k-1)x+\frac14(k+1)^2\tau-k\tau} \e{ x} f_S(S,\tau),
\label{eq:two-tilde-H_2-tmp}
\end{equation}
with zero initial condition. Hence
\begin{equation}
\tilde\H_2(\tau,x)=\frac2{\sigma^2}\int_0^\tau\int_{-\infty}^{\infty}\Psi(x-y,\tau-s)\e{\frac12(k-1)y+\frac14(k+1)^2s-ks}\e{ y} f_S(\e{y},T-\frac2{\sigma^2}s)dyds.
\label{eq:heat-eq-sol-tmp}
\end{equation}
From the boundedness of $Sf_S(S,t)$ it follows that $\tilde\H_2(\tau,x)=O(\e{\frac12(k-1)x})$, and we conclude that $SH_{2_S}^{\w}(S,t)$ is bounded on $\Rpplus\times\zeroClosedTclosed$. %
Using Remark \ref{remark:NT-width} it follows that $S^2Y^{\w} H_{2_S}^{\w}$ is bounded there too. Then $ H_3^{\wpm}(S,t)=\mp M^{\w}(T-t)\mp M_1^{\w}$, as defined in \eqref{eq:H_3-def}, will satisfy \eqref{eq:H_3} with $M^{\w}$ big enough and $M_1^{\w}$
as defined in \eqref{eq:M_1}. In either case, $SH_{3_S}^{\jp}(S,t)\equiv0,~j=1,w$.

$\hfill\Box$


\begin{thebibliography}{99}
{\rm

%
\bibitem{BarlesSoner}
{\sc Barles, G.\ \& Soner, H.~M.},
Option pricing with transaction costs and a nonlinear
Black-Scholes equation,
{\em Finance Stoch.} {\bf 2}, 369--397 (1998).
%
\bibitem{Bichuch}
{\sc Bichuch, M.} , Asymptotic analysis for optimal investment in finite time with transaction costs, {\em SIAM
J. Financial Math. } (To appear).
%
\bibitem{BichuchShreve}
{\sc Bichuch, M. \ \& Shreve, S.} , Utility maximization trading two futures with transaction costs,  preprint {\em www.math.cmu.edu/users/shreve/UtilityMaxOct30\_2011.pdf} (2011).

\bibitem{Bouchard}
{\sc Bouchard, B.}, Option pricing via utility maximization in the
presence of transaction costs: an asymptotic analysis, CEREMADE,
Univ.\ Paris Dauphine (1999).
%


%
\bibitem{BouchardTouzi}
{\sc Bouchard, B.\ \& Touzi, N.}, Explicit solution of the
multi-variable super-replication problem under transaction costs,
{\em Ann. Appl. Probab. }{\bf 10}, 685--708 (2000).



\bibitem{BouchardKabanovTouzi}
{\sc Bouchard, B., Kabanov Yu. M.\ \& Touzi, N.}, Option pricing by large risk aversion utility under transaction costs, {\em Decisions in Economics and Finance }{\bf 24}, 127--136 (2001).

%
\bibitem{BoyleVorst}
{\sc Boyle, P.~P.\ \& Vorst, T.},
Option replication in discrete time with transaction
costs,
{\em J.\ Finance} {\bf 47}, 272--293 (1992).



%
\bibitem{BurdzyKangRamanan}
{\sc Burdzy, K,~ Kang, W. \& Ramanan, K.},
The Skorokhod problem in a time-dependent interval, {\em Stochastic Processes and their Applications} {\bf 119}, 428Ð-452 (2009).
%
%

\bibitem{ClewlowHodges}
{\sc Clewlow, L.\ \& Hodges, S.D.}, Optimal delta-hedging under transactions costs, {\em J. of Econ.
Dynamics and Control.}
{\bf 21}, 1353--1376 (1997).

%
\bibitem{ConstantinidesZariphopoulou99}
{\sc Constantinides, G.\ \& Zariphopoulou, T.}, Bounds on prices
of contingent claims in an intertemporal economy with proportional
transaction costs and general preferences, {\em Finance Stoch.}
{\bf 3}, 345--369 (1999).
%
%
\bibitem{ConstantinidesZariphopoulou01}
{\sc Constantinides, G.\ \& Zariphopoulou, T.}, Bounds on
derivative prices in an intertemporal setting with proportional
transaction costs and multiple securities, {\em Math.\ Finance}
{\bf 11}, 331--346 (2001).
%

%
%
\bibitem{CEL} {\sc Crandall}, M. G., {\sc Evans}, L. C. and {\sc Lions}, P.-L., Some properties of viscosity solutions of Hamilton-Jacobi equations, {\em Trans. Amer.
Math. Soc.}\/  {\bf 282}, 487--502 (1984).
%
%
\bibitem{CrandallIshiiLions}
{\sc Crandall, M. G., Ishii, H.\ \& Lions P.L.}, User's guide to viscosity solutions of second order partial differential equations, {\em AMS Bulletin}
{\bf 1}, 1--67 (1992).
%
%
\bibitem{CL} {\sc Crandall}, M. G. and {\sc Lions}, P.-L., Viscosity
solutions of Hamilton-Jacobi equations, {\em Trans. Amer. Math. Soc.}\/ {\bf 277},
1--42 (1983).
%
%
%
%
%
%
%

%
\bibitem{CvitanicPhamTouzi}
{\sc Cvitani\v{c}, J., Pham, H.\ \& Touzi, N.},
A closed-form solution to the problem of
super-replicating under transaction costs,
{\em Finance Stoch.} {\bf 3}, 35--54 (1999).


%
\bibitem{CvitanicShreveSoner}
{\sc Shreve, S.\ E.\ \& Soner, H.\ M. and Cvitanic, J.},
There is no nontrivial hedging portfolio for option pricing with transaction costs, {\em Ann. Appl. Probab.} {\bf 5}, 327--355 (1995).
%
%
\bibitem {DavisNorman}
{\sc Davis, M.~H.~A.\ \& Norman, A.},
Portfolio  selection with transaction costs,
{\em Math.\ Oper.\ Res.} {\bf 15}, 676--713 (1990).
%

%
\bibitem{DavisPanasZariphopoulou}
{\sc Davis, M.~H.~A., Panas, V.\ G.\ \& Zariphopoulou, T.},
European option pricing with transaction costs, {\em SIAM J.\
Control} {\bf 31}, 470--493 (1993).
%
%
\bibitem{DelbaenKabanovValkeila}
{\sc Delbaen, F., Kabanov, Yu.\ \& Valkeila, E.},
Hedging under transaction costs in currency markets:
a discrete-time model,
{\em Math.\ Finance} {\bf 12}, 45--61 (2002).
%
%

 \bibitem{DumasLuciano}
 {\sc Dumas, B.\ \& Luciano, E.},
 An exact solution to a dynamic portfolio choice problem under
 transaction costs, {\em J.\ Finance} {\bf 46}, 577--595 (1991).




%
%
%
%
%
%
%
\bibitem{HodgesNeuberger}
{\sc Hodges, S.\ \& Neuberger, A.}, Option replication of
contingent claims under transaction costs, {\em Rev.\ Futures
Markets} {\bf 8}, 222--239 (1989).
%
\bibitem{JanecekShreve1}
{\sc Jane\v{c}ek, K.\ \& Shreve, S.}, Asymptotic analysis for optimal investment and consumption with transaction costs,
{\em Finance Stochastics} {\bf 8}, 181--206 (2004).
%
\bibitem{JanecekShreve2}
{\sc Jane\v{c}ek, K.\ \& Shreve, S.}, Futures 
trading with transaction costs, 
{\em Ill.\ J.\ Math.}, to appear.
%
%
\bibitem{KabanovLast}
{\sc Kabanov, Yu.\ \& Last, G.}, Hedging under transaction costs
in currency markets, {\em Math.\ Finance} {\bf 12}, 63--70 (2002).
%
%
\bibitem{KabanovStricker}
{\sc Kabanov, Yu.\ \& Stricker, C.} Hedging of contingent claims
under transaction costs. In: Sandman, K., Sch\"onbucher, P. (eds.)
{\em Advances in Finance and Stochastics. Essays in Honor of
Dieter Sondermann.} Springer 2002.
%

\bibitem{Karatzas}
{\sc Karatzas, I.}, Optimization Problems in the Theory of Continuous Trading,
 {\em SIAM J. Control} {\bf 27}, 1221--1259 (1989).


%
\bibitem{KoehlPhamTouzi97}
{\sc Koehl, P.~F., Pham, H.\ \& Touzi, N.}, Hedging in
discrete-time under transaction costs and continuous-time limit,
{\em J.\ Appl. Probab.} {\bf 36}, 163--178 (1999).
%
%
\bibitem{KoehlPhamTouzi98}
{\sc Koehl, P.~F., Pham, H.\ \& Touzi, N.}, On super-replication
under transaction costs in general discrete-time models, {\em
Theory of Probability and Applications} {\bf 45}, 783--788 (1999).
%
%
\bibitem{Leland}
{\sc Leland, H.},
Option pricing and replication with transaction costs,
{\em J.\ Finance} {\bf 40}, 1283--1301 (1985).
%
%
\bibitem{LeventhalSkorohod}
{\sc Leventhal, S.\ \& Skorohod, A.},
On the possibility of hedging options in the presence
of transaction costs,
{\em Ann.\ Appl.\ Probab.} {\bf 7}, 410--443 (1997).
%
\bibitem{Lott}
{\sc Lott, K.}, Ein Verfahren zur Replikation von Optionen unter Transaktionkostenin stetiger Zeit, {\em Dissertation}, Universit\"at der Bundeswehr M\"unchen, Institut fur Mathematik und Datenverarbeitung (1993).

%
%
\bibitem{MagillConstantinides}
{\sc Magill, M.~J.~P.\ and Constantinides, G.~M.},
Portfolio selection with transaction costs,
{\em J.\ Econ.\ Theory} {\bf 13}, 245--263 (1976).
%
%
\bibitem {Merton}
{\sc Merton, R.},
Optimum consumption and portfolio rules in a
continuous-time case,
{\em J.\ Econ.\ Theory} {\bf 3}, 373--413 (1971)
[Erratum {\bf 6}, 213--214 (1973)].
%
\bibitem {Pham}
{\sc Pham, H.},
Continuous-time stochastic control and optimization with financial applications,
{\em Springer-Verlag, Berlin} {\bf 61}, (2009).
%
%
%
%
%
%
%
%
%
%
%
%
%
%


%
\bibitem {ShreveSoner}
{\sc Shreve, S.\ \& Soner, H.\ M.}, Optimal investment and
consumption with transaction costs, {\em Ann.\ Applied Probab.}
{\bf 4}, 609--692 (1994).

%
\bibitem{SonerShreveCvitanic}
{\sc Soner, H.\ M., Shreve, S.\ \& Cvitani\v{c}, J.},
There is no nontrivial hedging portfolio for
option pricing with transaction costs,
{\em Ann.\ Appl.\ Probab.} {\bf 5}, 327--355 (1995).

%
\bibitem {WhalleyWilmott}
{\sc Whalley, A.~E.\ \& Wilmott, P.}, An asymptotic analysis of an
optimal hedging model for option pricing under transaction costs,
{\em Math.\ Finance} {\bf 7}, 307--324 (1997).
}
\end{thebibliography}
\end{document}